\documentclass{emulateapj}
\usepackage[utf8]{inputenc}
\usepackage{color}
\usepackage{natbib}
\usepackage[dvipsnames]{xcolor}
\usepackage{graphicx}
\newcommand{\kep}{{\em Kepler}}
\newcommand{\thisstar}{KIC 8462852}

\newcommand{\mearth}{{M$_\oplus$}}

\begin{document}

\title{KIC 8462852 Faded Throughout the \textit{KEPLER} Mission}

\newcommand{\caltech}{1}
\newcommand{\cfa}{2}
\newcommand{\carnegie}{3}

\author{%
Benjamin~T.~Montet\altaffilmark{\caltech,\cfa} and
Joshua~D.~Simon\altaffilmark{\carnegie}
}

\email{btm@astro.caltech.edu}

\altaffiltext{\caltech}  {Cahill Center for Astronomy and Astrophysics,
                          California Institute of Technology, Pasadena, CA,
                          91125, USA}
\altaffiltext{\cfa}      {Harvard-Smithsonian Center for Astrophysics,
                          Cambridge, MA 02138, USA}
\altaffiltext{\carnegie} {Observatories of the Carnegie Institution of Washington,
813 Santa Barbara St., Pasadena, CA 91101, USA}

\date{\today}

\begin{abstract}
\thisstar\ is a superficially ordinary
main sequence F star for which \kep\ detected an unusual series
of brief dimming events.  We obtain accurate relative photometry of \thisstar\ from the \kep\ full frame images, finding that the brightness of
\thisstar\ monotonically decreased over the four years it was observed
by \kep.  Over the first $\sim1000$~days, \thisstar\ faded
approximately linearly at a rate of $0.341 \pm 0.041\%$~yr$^{-1}$, for
a total decline of 0.9\%.  \thisstar\ then dimmed much more rapidly in
the next $\sim200$~days, with its flux dropping by more than 2\%.  For the final
$\sim200$~days of \kep\ photometry the magnitude remained
approximately constant, although the data are also consistent with the
decline rate measured for the first 2.7~yr.  
Of a sample of 193
nearby comparison stars and 355 stars
with similar stellar parameters, none exhibit the rapid decline by
$>2$\%\ or the cumulative fading by 3\%\ of \thisstar.
Moreover, of these comparison stars, only one changes brightness as quickly as
the
$0.341\%$~yr$^{-1}$ measured for \thisstar\ during the first three years of the
\kep\ mission.
We examine whether the
rapid decline could be caused by a cloud of transiting circumstellar material, finding
while such a cloud could evade detection in sub-mm observations, the 
transit ingress and duration cannot be explained by a simple cloud model.
Moreover, this model cannot account for the observed longer-term
dimming.
No known or proposed stellar phenomena can fully explain all aspects of the
observed light curve.
\end{abstract}
\keywords{stars: individual (\thisstar), stars: variables: general, circumstellar matter, techniques: photometric, methods: data analysis}

\maketitle

\section{Introduction}
\label{sec:intro}
In addition to its primary mission of detecting exoplanets \citep{Borucki10},
the \kep\ satellite's exquisite photometry has allowed groundbreaking
studies in stellar astrophysics.  Most notably, analyses of the
seismic modes of stellar lightcurves have enabled otherwise
inaccessible measurements of stellar ages, masses, evolutionary
states, and internal structure
\citep[e.g.,][]{Bedding11,Mosser12,Bastien13,Chaplin14,SilvaAguirre15}.
\kep\ data are also providing fundamental new insights into mass loss on
the red giant branch \citep[e.g.,][]{miglio12}, the variation of
stellar rotation and stellar dynamos with age \citep[e.g.,][]{Meibom15,Barnes16,vanSaders16}, and the origin of the Blazhko
effect in RR~Lyrae variables \citep[e.g.,][]{Szabo10}.

One of the most confounding \kep\ discoveries is the light curve
of \thisstar.
First presented by \citet{Boyajian16}, \thisstar\ appears to be a typical F3V star \citep{Lisse15}.
However, its light curve exhibits 10 significant dips over the timespan of the 
\kep\ mission. 
These dips are irregular in shape, aperiodic, and vary in depth from fractions of
a percent up to 20\% of the total flux of the star.  While the dips as a whole do not obey an obvious periodicity, a subset of them are consistent with a period of $\sim48$~d, although some of those are 180\degr\ out of phase with the others.
\citeauthor{Boyajian16} present several physical models to account for these dipping events,
concluding that a large family of exocomets or planetesimal fragments orbiting the
star could plausibly explain the data.

\thisstar\ quickly became the focus of considerable attention.
\citet{Thompson16} place tight constraints on the circumstellar dust around
\thisstar\ from millimeter and sub-millimeter observations, ruling out the
planetesimal fragment hypothesis;
\citet{Marengo15} reach the same conclusion from an analysis of warm \textit{Spitzer} data.
\citet{Bodman16} model the light curve with a swarm of comets.
They find that the \thisstar\ light curve in Quarters 16 and 17 of the \kep\ mission can be
explained by approximately 100 comets in a very tight cluster.
This model, however, is unable to match the earlier dips in the light curve.
\citet{Wright16} proposed that the star may make an ideal target for SETI 
programs.
As of yet, SETI searches in the optical \citep{Schuetz15, Abeysekara16} and
radio \citep{Harp15} have only resulted in null detections.

Other studies have made use of archival observations of \thisstar. 
\citet{Schaefer16} analyzed photographic plates from the 
``Digital Access to a Sky Century @ Harvard'' (DASCH) archive \citep{grindlay09,Laycock10}, obtained between 1890
and 1989.
This work found \thisstar\ to be dimming at an average rate of $0.165 \pm 0.013$~mag~century$^{-1}$, or $0.152 \pm 0.012\%$~yr$^{-1}$.
This result was quickly called into question.
Both \citet{Hippke16} and \citet{Lund16} suggest that the observed dimming is the result of
systematics in the DASCH data, particularly the Menzel Gap of the 1950s, and report the 
flux of \thisstar\ to be consistent with no change over the baseline of the DASCH plates. However, the DASCH team has not found evidence for such systematics \citep{Laycock10,Tang13a,Tang13b}.
Additionally, the claimed systematics appear to depend on the choice of reference stars (J. Grindlay 2016, priv. comm.).

If the star is dimming with time, then a survey with a long time baseline and high photometric
precision would be able to detect this variation.  
Modern ground-based imaging surveys generally operate over timescales
of $\sim5$~yr, and they lack the photometric accuracy to detect the
$\sim8$~mmag variation that would be expected from the analysis of the
DASCH data by \citet{Schaefer16}.  The best-calibrated existing
surveys, SDSS and Pan-STARRS \citep{Magnier13,Finkbeiner16}, would
need a time baseline $\sim5$ times as long in order to measure such a
brightness change at $5\sigma$ significance (ignoring that stars as
bright as \thisstar\ with $V = 11.7$ are badly saturated in such data sets).  Data from the
\kep\ mission, however, provide exactly such an opportunity.

The \kep\ mission was designed to detect and characterize transit events
with timescales of minutes to hours. 
To that end, \kep\ processing pipelines aim to remove longer-term trends in the data caused by instrumental
effects.
This process is well-described by \citet{Jenkins10} and \citet{Garcia11} and is the basis for the
creation of the ``pre-search data conditioning'' (PDC) light curves.
The PDC light curves, among many other corrections, remove long-term
trends in the observed flux of stars that correlate with either trends from nearby
stars or with the centroid of a star's motion during an observing quarter.
The main source of target motion is differential velocity aberration, 
which leads to an approximately linear drift in the centroid of a star at the 0.1 pixel level over the course of a quarter.
Any signals that are approximately linear over a quarter will then be removed as
an instrumental artifact.
Therefore, while the \citet{Boyajian16} analysis of \thisstar\ was sensitive to 
short-timescale variability, they or any other group
analyzing long cadence photometry of the target will not be able to detect any long-term
trends.
The end result is that signals with characteristic timescales longer than a single observing quarter will not appear in the long cadence data.
Searches for rotation periods in \kep\ data corroborate this
claim: \citet{Mcquillan14} identified more than 34,000 rotation
periods for stars in the \kep\ field, but their detection
efficiency drops for periods above 40 days and they do not
detect any rotation periods longer than 70 days.
Other sets of light curves designed to detect transiting
planets will more aggressively remove stellar variability
through a ``de-trending'' process, removing all non-transit signals \citep[e.g.][]{Carter12}.

The pixel-level data delivered by the telescope present a better opportunity to preserve long-term trends in the \kep\ light curves.
Photometry with the pixel-level data has its own difficulties. 
Aperture photometry is complicated as the apertures around each star recorded by the
telescope and downlinked to Earth are often smaller than the full point spread function (PSF) of
the star \citep{Bryson10}. 
Small deviations in the pointing of the telescope can then cause large variations in the
total flux recorded inside the photometric aperture as the edges of the star's PSF
move relative to the edge of the downloaded set of pixels.
Instead, one could consider modeling the PSF of the star on the detector.
While possible in some cases \citep[e.g.,][]{Rappaport14}, the lack of a reliable flat field for the
\kep\ detector array and the lack of background ``sky'' pixels make PSF modeling difficult, and nearly
impossible for saturated stars \citep[$K_p < 12$,][]{Szabo10}.

An alternative approach to search for long-term linear trends is to leverage the full-frame images (FFIs) collected during the \kep\ mission.
At the beginning of the mission, \kep\ recorded a series of eight FFIs over two days.
It then continued to record one FFI each month throughout the mission \citep{Haas10}.
As \kep\ rotated by 90\degr\ every three months, there are generally three consecutive FFIs with all stars located on the same pixels.  A star then cycles through four different detectors over the course of a year, returning to its original position and repeating the cycle a year later.  Because there are $\sim13$ observations with any given star landing on the same pixels, accurate relative photometry can be obtained without flat fielding.  The FFIs have been used to study stellar variability in RR Lyrae stars in the \kep\ field \citep{Kinemuchi11}, highlighting their utility for photometry.

The primary \kep\ mission lasted four years. A star dimming at a linear rate of 0.165 mag
century$^{-1}$ ($0.152 \pm 0.012\%$~yr$^{-1}$), as has been purported for \thisstar\ \citep{Schaefer16}, would be expected to decrease in brightness by 0.6\% over the \kep\ baseline, well
above the photometric precision of the telescope.
By analyzing the FFI data, one could avoid the removal of this trend by the data processing
pipeline and the loss of flux from limited apertures in the raw pixel-level data.
In this study, we attempt to characterize the long-term photometric behavior of \thisstar\
through an analysis of \kep\ FFI data.

This paper is organized as follows.
In Section \ref{sec:data}, we describe our data analysis and photometry.
In Section \ref{sec:results}, we present our results.
In Section \ref{sec:compare}, we compare our results to a similar analysis of stars
close to \thisstar\ on the detector and stars with similar stellar properties.
In Section \ref{sec:discuss}, we rule out instrumental effects that could be responsible for apparent brightness changes and consider whether background contamination, a transiting cloud of material, or a polar spot can explain the observed FFI light curve.  We present our conclusions in Section~\ref{sec:conclusions}.

\section{Data Analysis}
\label{sec:data}

\begin{figure}[htbp!]
\centerline{\includegraphics[width=0.5\textwidth, trim={1cm 1cm 1cm 1cm},clip]{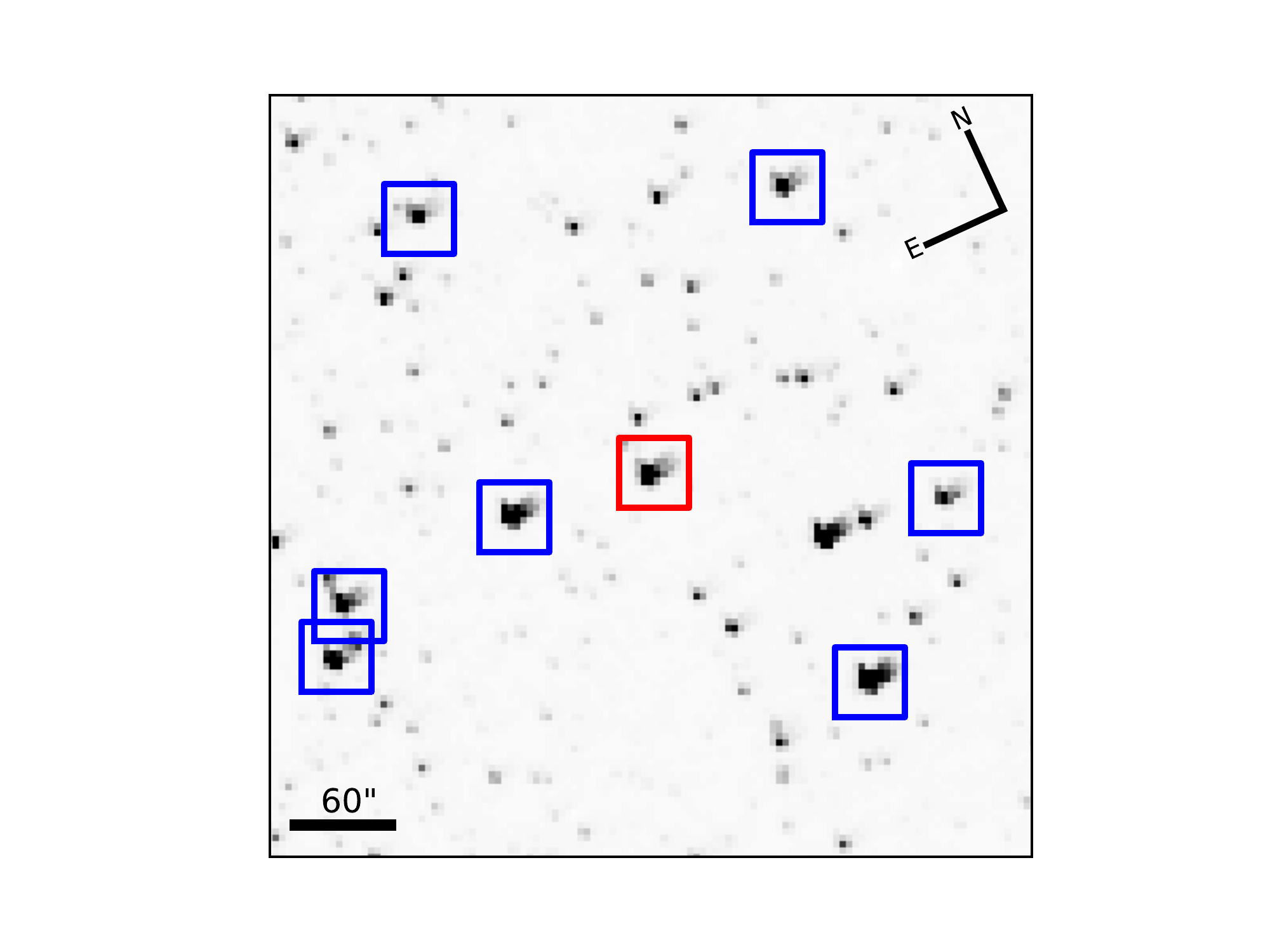}}
\caption{``Postcard'' region of the \kep\ detector surrounding \thisstar, located at the center
of the image. 
In red is the aperture used for measuring the photometry from this star in all frames.
The seven blue apertures represent the reference stars used for comparison photometry.
The frame is aligned to match the orientation of the telescope, with North
rotated approximately 25 degrees from pointing up.
Most of the stars visible in this image were not observed in long or short cadence
during the \kep\ mission; the FFI data is the only photometry available for these targets.
}
\label{fig:FFI}
\end{figure}

We downloaded all 53 FFI images from the Mikulski Archive for Space Telescopes (MAST) and sorted
by the season in which they were collected.
In this work, we use the ``cal'' frames, which have been calibrated from the raw data 
\citep{Caldwell10, Jenkins10, Quintana10}.
Specifically, these frames have had a bias level and dark current subtracted, the smear from CCDs being illuminated during 
readout corrected, the gain divided, and a flat field correction applied.
We removed one frame, 2009170043915, as it was collected with the spacecraft mispointed by four pixels on the
sky, making calibration with the other frames in that spacecraft orientation infeasible.
Fluxes from the 2009170043915 frame differ from the other
frames in that particular orientation taken during that same quarter by
0.5\%, suggesting that the offset in the flat field from pixel to pixel is at the
sub-percent level.
For the remaining 52 frames, we first select a $120 \times 120$ pixel (8 arcmin square) region of the detector centered on the
target star (Figure \ref{fig:FFI}).
As a part of the FFI data processing, a background has already been subtracted.
Unlike in the \textit{K2} mission, which observes in the ecliptic and detects a changing
background due to increasing zodiacal light during each campaign \citep{Molnar15}, 
the background level is low and does not significantly change during the \kep\ mission.

\begin{figure}[htbp!]
\centerline{\includegraphics[width=0.50\textwidth, trim={1cm 1cm 1cm 1cm},clip]{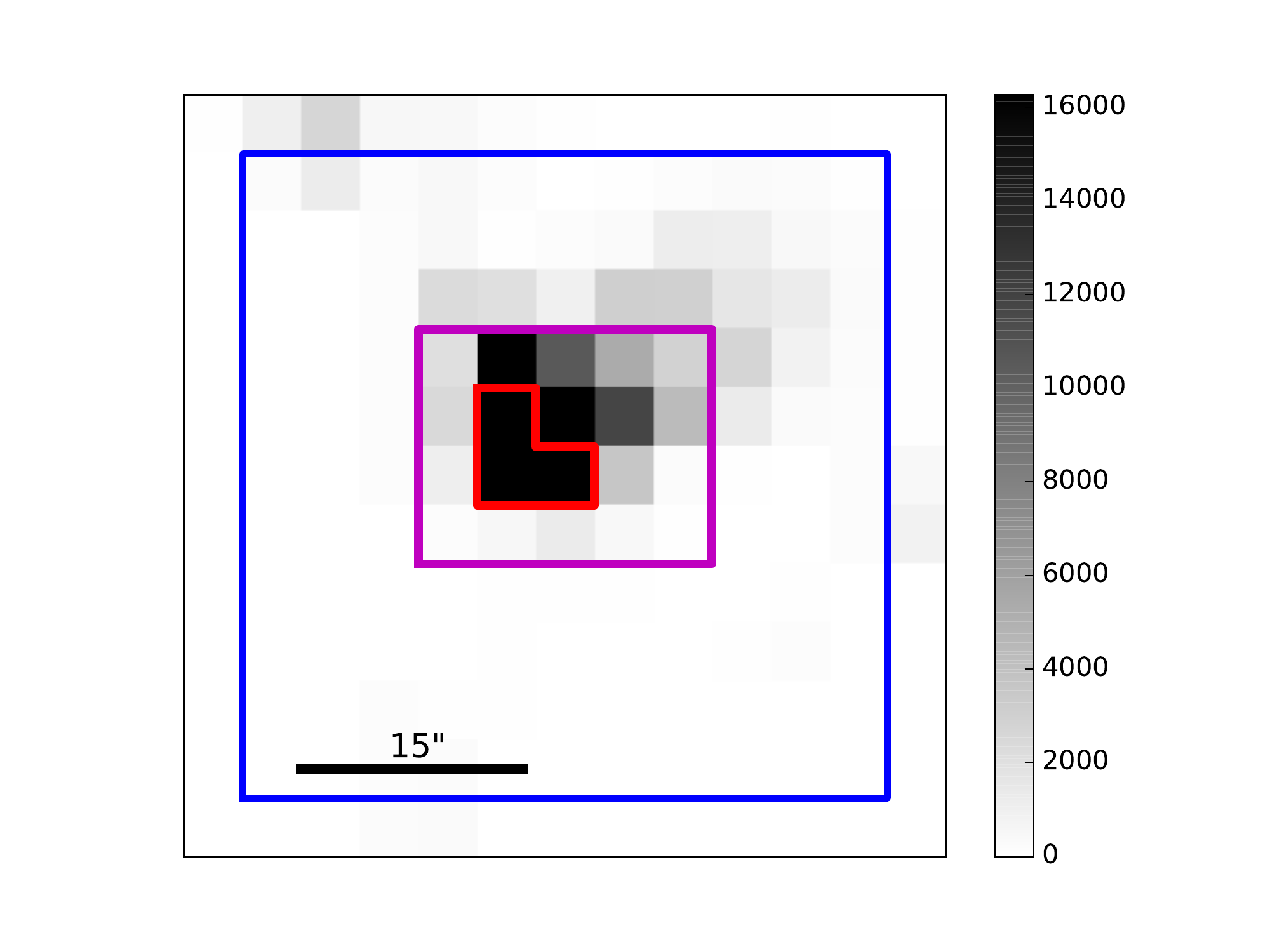}}
\caption{Zoomed region of Figure \ref{fig:FFI} centered on \thisstar.
In blue is the aperture used for measuring the photometry from this star in all frames.
In purple is the postage stamp downloaded for the creation of the long cadence light curve 
for this star, while in red are the actual pixels used in the creation of the SAP and PDC 
light curves in Quarter 0 of the mission.
The pixel response function is wider than the regions downloaded at 30-minute cadence, so not
all of the flux from \thisstar\ is captured.
The orientation here is the same as in Figure 1.
}
\label{fig:lFFI}
\end{figure}

We begin our processing by masking any saturated stars.
While \kep\ was designed with preservation of photometry of saturated stars in mind, recovering the absolute brightness of these
stars requires careful placement of photometric apertures, making automated procedures challenging.
We then identify the ten brightest nearby stars in the subframes based on their brightest central pixel.

For all selected stars, including our target, we perform aperture photometry. 
The \kep\ telescope is defocused to produce an image with a full-width at half-maximum of 1.5 pixels in diameter.
However, due to the location of \thisstar\ in the corner of the \kep\ field, 
the Schmidt optics of the telescope produce a non-Gaussian and asymmetric PSF for each star \citep{Bryson10}.
As can be seen in Figure \ref{fig:lFFI}, these PSFs are several pixels
in size and extend well beyond the pixels downloaded at 30-minute cadence 
during standard \kep\ observations.
To collect all the flux from the star, we then create $11\times11$ pixel apertures
centered on the brightest pixel of each star. 

Not all of these stars may be appropriate as calibration stars for relative photometry; some may be intrinsically variable, while
others may be too faint to reliably obtain accurate photometry in each single exposure.
We discard all stars with a standard deviation of the residuals from a linear fit to the data larger than 0.5\%.
Such a cut removed one bright star near \thisstar\ (KIC 8462738) with periodic variations at the $>5$\% level 
as well as the two faintest stars in the sample, where photon noise
prohibited accurate relative photometry, leaving us with seven reference stars.

We then determine the flux level of \thisstar\ by comparing its aperture
photometry to the aperture photometry recorded from the calibrator stars, weighted by the signal to noise ratio of each of the calibration
stars.
We use the standard deviation of the flux measured among all observations recorded on that particular pixel as the uncertainty associated with each observation.
The scatter between points on a particular detector can vary by as much as a factor of two from detector to detector, in
line with previous analyses of the noise in the primary \kep\ mission 
\citep{Gilliland11}.

\begin{figure*}[htbp!]
\centerline{\includegraphics[width=0.9\textwidth]{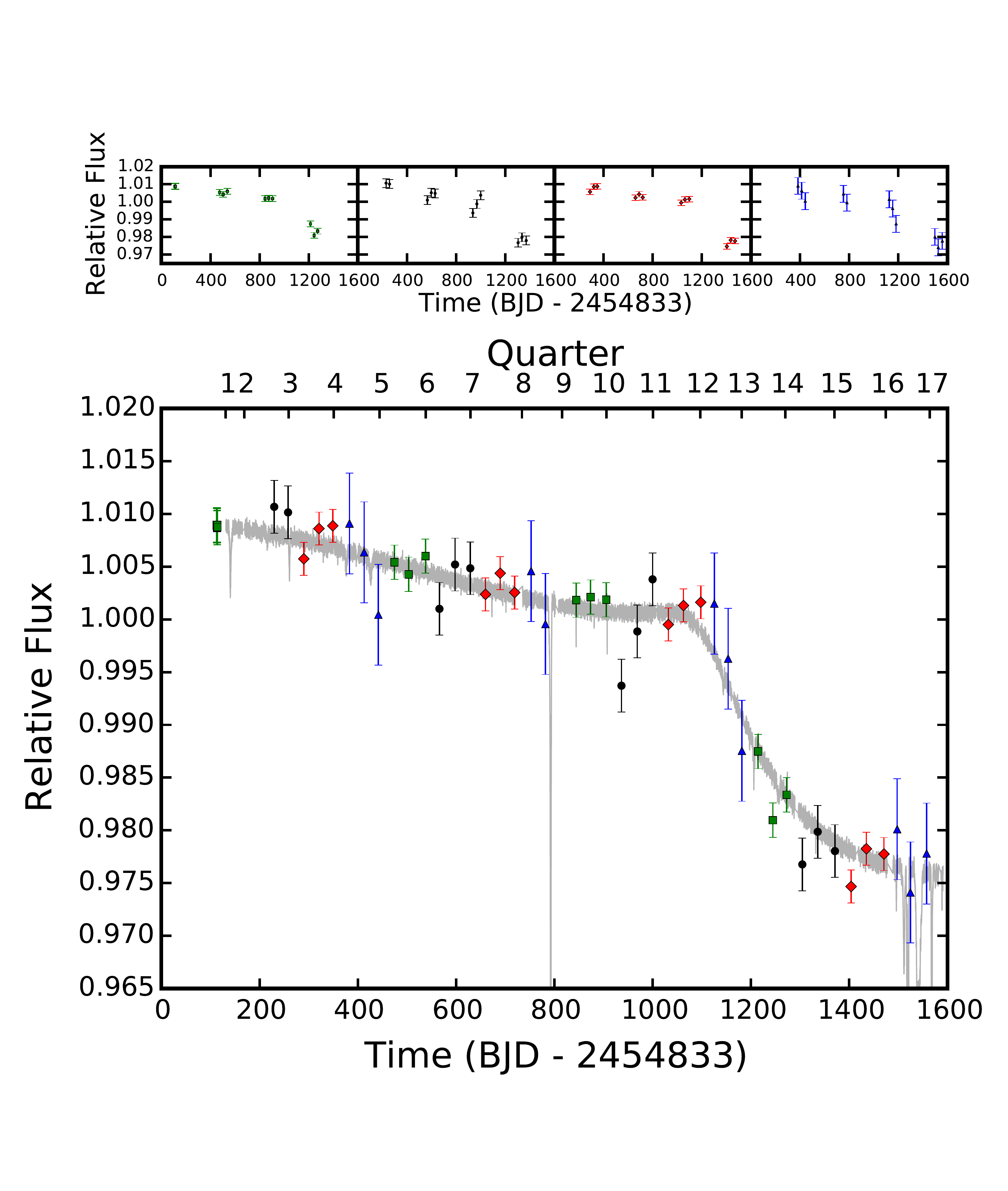}}
\caption{Photometry of \thisstar\ as measured from the FFI data. The four colors and shapes
(green squares, black circles, red diamonds, and blue triangles) represent measurements
from the four separate channels the starlight reaches as the telescope rolls. The four subpanels show the flux measurements from each particular detector 
individually. The main figure combines all observations together.
In the process of creating the fit, we allow a vertical offset between the
data from each individual quarter to account for changes in the flat field with
detector orientation. For the purposes of this figure, we plot the maximum likelihood values: the blue, green, and black points have been shifted upward
by 0.6\%, 0.1\%, and 0.2\%, respectively, suggesting variations in the flat
field between pixels are at the sub-percent level.
In all four channels, the photometry is consistent with a linear decrease in flux for the
first three years of the mission, followed by a rapid decrease in flux of $\approx 2.5\%$.
The light gray curve represents one possible \kep\ long cadence light curve consistent with the 
FFI photometry created by fitting a spline to the FFI photometry as described in Section 4. The large
dips observed by \citet{Boyajian16} are visible but narrow relative to the cadence of FFI observations.
The long cadence data are available as Data Behind the Figure, available in the online version of the journal.
}
\label{fig:photometry}
\end{figure*}

Finally, we combine the data from each separate channel on the telescope.
Because of the uncertainty in the underlying flat field, 
we expect an offset between each telescope orientation
which is not known \textit{a priori}, qualitatively similar to those observed
in the \kep\ long cadence data between quarters.
We first normalize the flux values as recorded in the previous 
paragraph by dividing the observations from each particular channel
by their mean value.
We then apply a
linear offset term between each channel, so that if the sensitivity of
one particular channel is lower than the others, our model would not
try to model that signal is actual astrophysical variation.
In practice, this requires fitting three parameters defining the 
 offset for each channel relative to the first channel.
The results are shown in Figure \ref{fig:photometry} and Table 1.
For display purposes in Figure \ref{fig:photometry} we must select a
particular set of offsets.
We choose the maximum likelihood model of a linear fit to the first three
years of data from the \kep\ mission, which minimizes the scatter from quarter to quarter.
In our analysis of any long-term dimming, we allow the offset terms to float as free parameters.

\begin{deluxetable*}{ccccccc}
\tablecaption{FFI Data}
\footnotesize
\tablewidth{0pt}
\tablehead{
  \colhead{Time} & 
  \colhead{Observed Flux} &
  \colhead{Reference Flux\tablenotemark{1}} &
  \colhead{Flux Ratio} &
  \colhead{Normalized Flux\tablenotemark{2}}     &
  \colhead{Uncertainty} &
  \colhead{Orientation}  \\
  \colhead{(BJD - 2454833)} & 
  \colhead{(e s$^{-1}$)} &
  \colhead{(e s$^{-1}$)} &
  \colhead{} &
  \colhead{} &
  \colhead{($1\sigma$)} &
  \colhead{}      
}
\startdata
112.742 & 277998 & 1384241 & 0.200831 & 1.00896 & 0.00163 & 3 \\
112.867 & 278052 & 1384642 & 0.200811 & 1.00886 & 0.00157 & 3 \\
113.018 & 278029 & 1384472 & 0.200820 & 1.00890 & 0.00158 & 3 \\
113.234 & 277993 & 1384621 & 0.200772 & 1.00867 & 0.00161 & 3 \\
113.338 & 278018 & 1384317 & 0.200834 & 1.00898 & 0.00161 & 3 \\
113.550 & 278053 & 1384618 & 0.200816 & 1.00888 & 0.00163 & 3 \\
113.733 & 278046 & 1384486 & 0.200830 & 1.00895 & 0.00162 & 3 \\
114.166 & 277968 & 1384404 & 0.200785 & 1.00873 & 0.00165 & 3 \\
229.825 & 283733 & 1408715 & 0.201413 & 1.01067 & 0.00257 & 0 \\
258.006 & 283423 & 1407913 & 0.201307 & 1.01014 & 0.00267 & 0 \\
290.086 & 268042 & 1332411 & 0.201171 & 1.00574 & 0.00153 & 1 \\
320.980 & 268832 & 1332528 & 0.201746 & 1.00861 & 0.00162 & 1 \\
349.037 & 268720 & 1331627 & 0.201798 & 1.00887 & 0.00156 & 1 \\
382.955 & 275452 & 1377897 & 0.199908 & 1.00908 & 0.00484 & 2 \\
383.035 & 275495 & 1378097 & 0.199910 & 1.00909 & 0.00483 & 2 \\
412.766 & 275353 & 1381124 & 0.199369 & 1.00636 & 0.00460 & 2 \\
441.740 & 273902 & 1381975 & 0.198196 & 1.00044 & 0.00502 & 2 \\
474.535 & 276125 & 1379789 & 0.200121 & 1.00541 & 0.00156 & 3 \\
503.428 & 275547 & 1378483 & 0.199891 & 1.00427 & 0.00154 & 3 \\
537.695 & 275378 & 1375253 & 0.200238 & 1.00600 & 0.00173 & 3 \\
566.057 & 279781 & 1402980 & 0.199419 & 1.00100 & 0.00256 & 0 \\
597.811 & 280641 & 1400955 & 0.200321 & 1.00520 & 0.00248 & 0 \\
628.829 & 280421 & 1400350 & 0.200251 & 1.00485 & 0.00266 & 0 \\
659.806 & 265617 & 1324798 & 0.200496 & 1.00238 & 0.00153 & 1 \\
689.762 & 266120 & 1324664 & 0.200896 & 1.00437 & 0.00157 & 1 \\
719.084 & 265682 & 1324891 & 0.200531 & 1.00255 & 0.00158 & 1 \\
752.576 & 272275 & 1368110 & 0.199015 & 1.00458 & 0.00483 & 2 \\
781.739 & 271728 & 1372220 & 0.198021 & 0.99956 & 0.00462 & 2 \\
844.444 & 273245 & 1370321 & 0.199402 & 1.00183 & 0.00166 & 3 \\
873.644 & 272944 & 1368418 & 0.199922 & 1.00211 & 0.00163 & 3 \\
905.459 & 272206 & 1365061 & 0.199409 & 1.00186 & 0.00159 & 3 \\
936.477 & 275546 & 1391437 & 0.198030 & 0.99371 & 0.00247 & 0 \\
968.762 & 276586 & 1389492 & 0.199055 & 0.99885 & 0.00236 & 0 \\
999.801 & 277627 & 1387847 & 0.200415 & 1.00380 & 0.00248 & 0 \\
1031.800 & 263255 & 1316791 & 0.199922 & 0.99951 & 0.00153 & 1 \\
1062.757 & 263785 & 1317059 & 0.200283 & 1.00132 & 0.00151 & 1 \\
1097.862 & 263738 & 1316417 & 0.200345 & 1.00162 & 0.00156 & 1 \\
1125.427 & 269393 & 1357792 & 0.198405 & 1.00150 & 0.00453 & 2 \\
1153.523 & 268327 & 1359507 & 0.197371 & 0.99628 & 0.00474 & 2 \\
1181.558 & 266378 & 1361569 & 0.195641 & 0.98754 & 0.00505 & 2 \\
1214.517 & 266884 & 1357982 & 0.196530 & 0.98748 & 0.00168 & 3 \\
1244.453 & 264824 & 1356515 & 0.195224 & 0.98096 & 0.00166 & 3 \\
1272.590 & 264792 & 1353002 & 0.195707 & 0.98337 & 0.00166 & 3 \\
1304.527 & 268808 & 1380987 & 0.194649 & 0.97677 & 0.00250 & 0 \\
1335.832 & 269135 & 1378297 & 0.195266 & 0.97986 & 0.00237 & 0 \\
1370.855 & 268435 & 1377283 & 0.194902 & 0.97803 & 0.00256 & 0 \\
1403.835 & 255025 & 1308199 & 0.194944 & 0.97467 & 0.00168 & 1 \\
1434.914 & 255839 & 1307564 & 0.195661 & 0.97825 & 0.00150 & 1 \\
1470.673 & 255686 & 1307448 & 0.195561 & 0.97775 & 0.00156 & 1 \\
1497.564 & 261275 & 1345623 & 0.194167 & 0.98010 & 0.00469 & 2 \\
1524.495 & 260101 & 1347812 & 0.192980 & 0.97411 & 0.00456 & 2 \\
1557.495 & 261323 & 1349036 & 0.193711 & 0.97780 & 0.00496 & 2 
\enddata
\tablenotetext{1}{The ``Reference Flux'' is the sum of the
fluxes of the seven reference stars, weighted by the signal-to-noise
ratio of each star on the detector.}
\tablenotetext{2}{The ``Normalized Flux'' is the Flux Ratio
for each star normalized by quarter as described in 
Section 2.}
\label{tab:results}
\end{deluxetable*}

\section{Results}
\label{sec:results}

\subsection{A Long-Term Dimming}
\label{sec:longtermdimming}

From Figure \ref{fig:photometry}, three main features are apparent.
The first is a long-term dimming at a linear rate from the beginning of the \kep\
mission until approximately Quarter 13.
The dimming then increases rapidly such that the observed flux from the
star decreases by approximately 2.5\%\ over 200 days. 
For the final 200 days of the \kep\ mission, the light curve flattens out, either
returning to the original rate of decay or remaining constant.

In the primary \kep\ mission, data were collected in four orientations as the telescope
rolled every 93 days to keep its heat shield pointed at the sun.
As a result, data for each star are collected on four different pixels, each for three months
of every year. 
Therefore, we can separate many instrumental effects caused by a faulty pixel on the detector
from astrophysical phenomena by looking for the same trends when only considering data from
each specific orientation.
In practice, we allow for a linear offset between the observed fluxes recorded on each
individual module, reflecting our uncertainty about the underlying flat field of the 
detector itself.
We note that these results are apparent in data from each individual detector, not just the
combined light curve, suggesting that the decline in flux is an astrophysical effect rather than an 
instrumental one, as 
discussed more fully in Section \ref{sec:discuss}.

We fit a line to the first region of data, from the beginning of the mission until
Barycentric Julian Date (BJD) 2456003
(or BJD-$2454833 = 1170$), encompassing 39 data points.
We allow the relative offset between each detector to vary, three in total to account for possible differences in the flat field between different detectors.
It is possible that the recorded measurement uncertainties are not representative of the actual uncertainties.
To account for a possible systematic underestimation of the uncertainties,
we allow for an extra
uncertainty term applied to data from each detector to be added in quadrature to our
recorded uncertainties estimated from the data themselves, so that
the $i$th data point recorded in the $q$th telescope orientation has uncertainty 
$s_{i,q}$ such that
\begin{equation}
    s_{i,q} = \sqrt{\sigma_{i}^2 + j_q^2}.
\end{equation}
Here, $\sigma_i$ is the uncertainty for the $i$th observation (as described in \S\ref{sec:data}) and
$j_q$ is the level of underestimation of the photometric uncertainty in 
quarter $q$, similar to the concept of jitter in radial velocity (RV) 
observations \citep{Butler96b}. 
We assume all observations on a given detector have the same $j_q$ value, so that
there are four $j_q$ values fit in our analysis.

Including a slope and zeropoint for the linear fit, there are a total of nine free parameters: the slope, zeropoint, three offset terms describing the linear
offset for each quarter relative to the first, and four ``jitter'' terms describing
the excess photometric noise in each channel above the 
listed uncertainties.
The resultant fit estimates the excess noise required in order 
to explain the data for the first three years of the mission as a straight line, and
estimates the offset in flux between the different spacecraft orientations under
the assumption that the rate of change of flux is constant over the first three
years of the mission.
We note that the data shown in Figure 3 and Table 1 are the uncertainties from the
data themselves ($\sigma_i$) rather than $s_{i,q}$.

By marginalizing over all parameters except the slope, we can measure the decay rate
of the observed flux.
We determine that the star is fading from our perspective at a rate of $0.341 \pm 0.041\%$ yr$^{-1}$, or $3.41 \pm 0.41$ parts per thousand (ppt) yr$^{-1}$.
This is equivalent to fading at a rate of $0.370 \pm 0.044$ magnitudes per century, exceeding the purported 0.165 mag
century$^{-1}$ ($0.152 \pm 0.012\%$~yr$^{-1}$) rate of dimming over the
duration of the DASCH plates \citep{Schaefer16} by more than a factor of two.  The total change in brightness over this portion of the \kep\ light curve is almost 1\%.
The four maximum likelihood ``jitter'' values are 
2.8, 0.52, 1.1, and 0.28 parts per thousand, corresponding to orientations 0, 1, 2, and 3 respectively in the parlance of Table 1, suggesting that the recorded
uncertainties listed there are not significantly underestimated.

\subsection{A Rapid Dimming}

Beginning around or BJD$-2454833 \sim 1100$, lasting for approximately 200 days, the
rate of dimming increases dramatically. Repeating the above exercise in the region between dates 1100 and 1250, we measure a decay rate of $3.37 \pm 1.48\%$ yr$^{-1}$. Here, we allow the break points at which the slope changes to float as
free parameters, and the uncertainty in the actual dimming rate during this event is dominated by the uncertainty in the exact timing of the change in slope. 
Regardless of the exact rate at which the star appears to dim from our line of sight,
it is clear that from the \kep\ data, the star becomes approximately 2.5\% dimmer over a
period of 200 days.
This rapid dimming is apparent in data from all four orientations of the \kep\ spacecraft.
We note that we would expect the \kep\ pre-search data conditioning (PDC) pipeline to remove
signals such as this one in the pre-processing of the data, explaining why this signal does
not appear in the originally published long-cadence light curve.

After the conclusion of this event, there are only nine epochs of FFI data.
Over this range, the rate of change of flux from the star is poorly measured: it is 
consistent with the original rate of dimming, but also consistent with a flat light curve until the final FFI image is collected at the end of Quarter 16 of the \kep\ mission.

We offer no definitive explanation that could explain the observed light curve in this work. 
The effect could be stellar in nature, although there are no known mechanisms that
would cause a main-sequence F star to dim in brightness by 2.5\% over a few months.
The effect could also be caused by a passing dust cloud in orbit around \thisstar.
Indeed, the light curve at times larger than 1000 days has a morphology broadly
similar to a transit event, although on markedly different timescales. 
We discuss this morphology more fully in Section \ref{sec:transit}.

\subsection{Comparison with Dimming Events}
\label{sec:cwde}

\begin{figure}[htbp!]
\centerline{\includegraphics[width=0.45\textwidth]{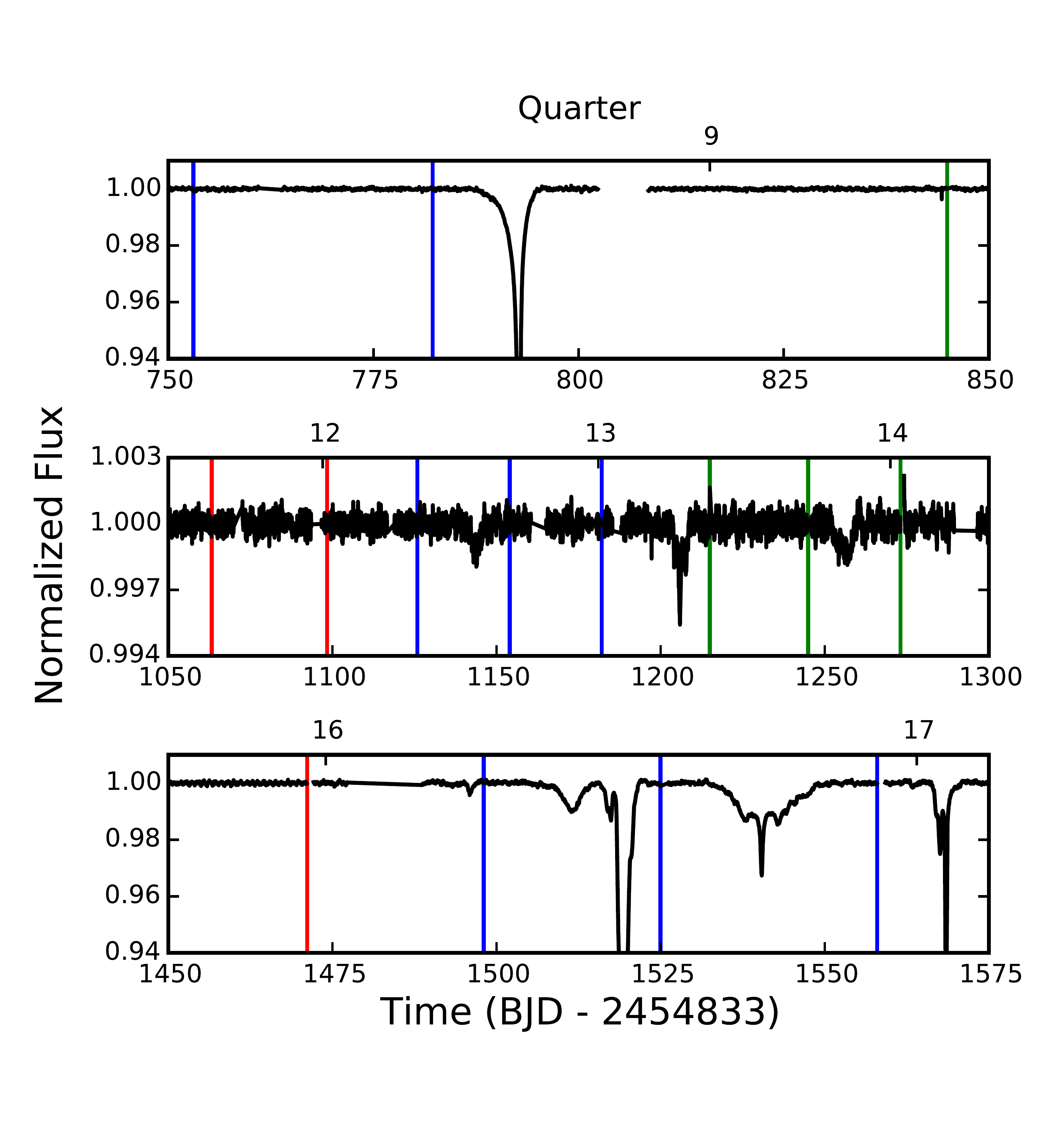}}
\caption{Times of FFI data collection superimposed on the \kep\ long cadence light curve
of \thisstar. Each vertical line represents the time of capture of an FFI, with the colors
retaining their same meaning as in Figure \ref{fig:photometry}. 
None of the FFIs are taken during a dip that could significantly affect our photometry.
We produce our own long cadence light curve as explained in Section 5.3,
finding no evidence for the rise before the dip at the end of Quarter 8 observed in the 
PDC light curve. The light curve in this figure is available as the Data
Behind the Figure associated with Figure 3 and available in the online version of this journal.
}
\label{fig:lc}
\end{figure}

The long-cadence light curve as observed by \kep\ contains features in which the
observed flux from the star decreases by as much as 20\% for a few days at a time.
Our results could be substantially affected if any of these events were to overlap with the collection of a full frame image.
We can test this possibility by comparing the times recorded for the FFI data with the largest
dipping events in the \kep\ light curve. 
The result is shown in Figure \ref{fig:lc}.

We find that the dips do not overlap with any of the FFI observations. 
The only FFI image that is sufficiently close to a dipping event to possibly
be affected is the final FFI of Quarter 8. 
If we repeat our analysis removing this data point, we recover the same long-term
dimming; none of the results in this work are significantly affected by the inclusion
or exclusion of this single data point.

FFI images are typically collected as the last observation of a month of continuous data
collection immediately before data downlink to Earth.
After the FFI is taken, there is typically a $\approx$1 day gap before data collection
is resumed.
It is unlikely that a large dip could fall in one of these gaps while evading detection
in the long cadence data, as the ingress duration for a dip would need to be smaller 
than the 30 minutes between the last long cadence image of a single month and the FFI 
collected immediately afterward.

\subsection{Comparison with Rising Events}

As can be seen in Figure~1b of \citet{Boyajian16}, some of the dipping events are surrounded by apparent flux increases of $\sim0.1\%$.  If these increases are real, they could be the signature of forward scattering from dust grains \citep[e.g.,][]{Rappaport12}.  Most notably, the PDC light curve produced by the \kep\ mission appears to rise just before a 15\% dip.
This particular
FFI frame records a lower flux value than its predecessor by approximately 1 mmag,
although the two are consistent with no change in flux. 
To test whether the small rise in the PDC light curve is induced by the 
initial data processing in the creation of the PDC light curves or if it is real,
we create our own long cadence light curve from the pixel-level data.

We perform aperture photometry using the entire postage stamp downloaded by the \kep\
mission, not just the smaller aperture selected by the \kep\ team for the creation of
SAP and PDC light curves.
We do the same for KIC 8462934, the single star within 2 arcmin with $K_p$ within one magnitude
of \thisstar.
We then divide the flux recorded for \thisstar\ by the flux recorded for its apparent neighbor
to account for instrumental trends.

In the full light curves, 
we detect a significant variation in the total flux received from quarter to quarter, a
sign of changes in the underlying flat field from channel to channel.
For each quarter, we fit an offset term to minimize the variation between quarters
(see \S\ref{sec:longtermdimming}). All light curves produced from \kep\ data contain
long-term variability that can result from a combination of astrophysical and instrumental effects. With the FFI data, we are able to use the nearby stars on the detector to separate
the astrophysical effects from any instrumental effects shared by nearby stars.
However, as the vast majority of stars in the Kepler Input Catalog were not targeted
for regular photometry during the \kep\ mission, there is a paucity of stars
with long cadence data to compare against to robustly detect long-term astrophysical
trends in the long-cadence data.
In this case, for the purposes of plotting Figure 4 we remove long-term trends.
We divide the light curve into distinct regions separated by
gaps in the data of at least 0.5 days.
We then mask the regions of the light curve identified as short-term 
dips by \citet{Boyajian12} and fit a spline to the remainder of the
data, dividing the measured light curve by the spline to flatten the
light curve.
The result is a normalized light curve that removes
trends but preserves the shape and magnitude of the rapid dips.
These data are shown in Figure \ref{fig:lc} and we make them publicly available as ``Data Behind the Figure'' associated with Figure 3.

Once we have re-processed the light curve, we can compare the rising events to 
those observed in the PDC light curve.
The rise in the PDC light curve in Q8 does not exist in our re-processed light curve, suggesting that it is
an artifact of the processing in the development of the PDC light curve.
Similarly, the rise in the light curve at Day 1148, observed in the PDC light curve and
this light curve, is almost certainly an artifact of data processing.
There are also small rises observed before the dips at dates 140 and 260.
We note that these also occur immediately after data downlinks, as the
telescope is reestablishing thermal balance after rotating to point to earth and changing the position of its heat shields.
It is well established that this maneuver leads to a ramp in the observed
flux lasting $\approx 3$ days after observations resume \citep{Smith12}.
As both rises occur during these windows, it is likely that they are
induced by these thermal effects.
There is a small rise ($\approx 2$ mmag) observed at the end of Quarter 13 that may be real; it appears at low significance in the FFI data as well. 

While our re-processed light curve is useful for studying the short-term
dimming and potential rising events in the light curve of \citet{Boyajian16}, it and other versions of the 
\kep\ long cadence data for this star are not well-suited for the detection of long-term variations in the observed
flux (see \S\ref{sec:intro}). 
The total aperture recorded by \kep\ at 30-minute cadence is smaller than the PSF of the
detector, so small changes in the position of the star cause flux variations that easily exceed the magnitude of the long-term trend.
Moreover, many of the nearby reference stars used in this work that should share systematics with \thisstar\ were not observed at  
30-minute cadence, meaning the opportunity to co-trend systematics as robustly as done here
is lost. 
The best opportunity to create a light curve that is accurate both on traditional \kep\
timescales and over years is to combine the FFI data with long cadence data, as we discuss in Section 5.3.
This light curve can also be used to test the robustness of some of the signals in the PDC
light curve.

\section{Comparison to Other \textit{KEPLER} Stars}
\label{sec:compare}

The long-term dimming of \thisstar\ observed during the first 12 quarters of the \kep\ mission differs from zero by 
$8\sigma$.
In modeling the light curve, we assumed that the sources of uncertainty are entirely
statistical in nature.
If there are other, long-term systematics that cause a slow drift in the
recorded \kep\ flux from a star, those could evade our statistical analysis and cause us to
erroneously record a long-term trend that does not exist.
If such systematics exist, their effect should be detectable for many stars, so we can compare the
observed light curve in our analysis with other stars in the \kep\ field to see
how often a star exhibits a dimming of $0.341 \pm 0.041\%$ yr$^{-1}$. 

\subsection{Reference Stars}

We repeat the analysis in Section \ref{sec:results}, performing photometry on other stars that fall onto
the $8\times8$ arcminute postcard around
\thisstar.
If long-term trends of a similar magnitude are recovered in any of these, it is an 
indication that the \kep\ photometry is dominated by systematics from one or a few
active stars in particular.

\begin{figure}[htbp!]
\centerline{\includegraphics[width=0.45\textwidth]{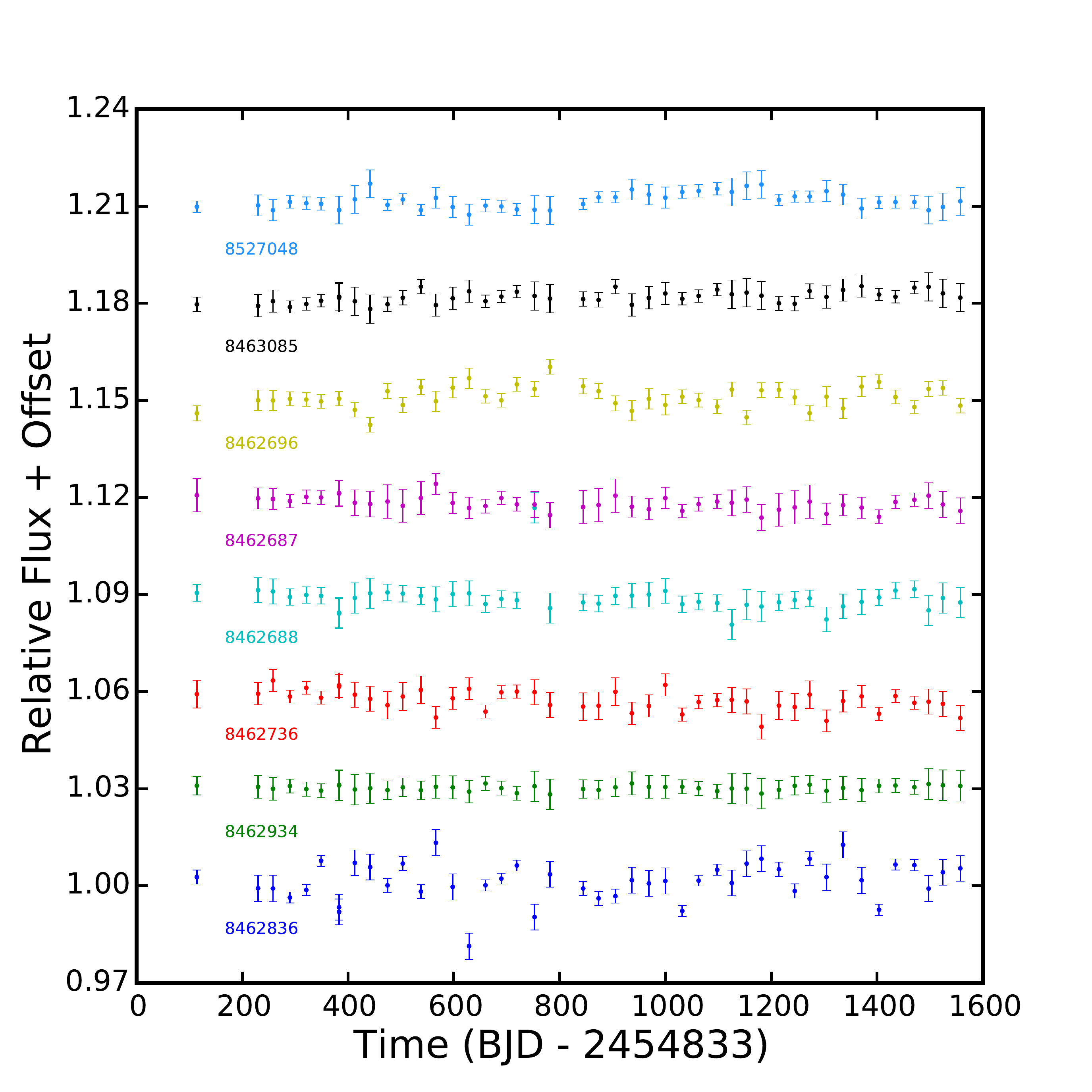}}
\caption{Photometry for eight stars produced in the same method as described in Section 2.
These stars all have $11 < K_p < 14.5$ and are located within $4'$ of \thisstar\ on 
the sky. 
In no cases do we detect a similar dimming to that observed for \thisstar, suggesting that the
dimming is not caused by systematics due to variability of nearby reference stars.
}
\label{fig:refs}
\end{figure}

To test for such systematics, we repeat our process to develop light curves for eight
other stars, all with $11 < K_p < 14.5$ and all located within $4'$ (60 pixels) of
\thisstar\ on the sky.
In these cases, \thisstar\ was not selected by our pipeline as a potential reference
star due to its photometric variability relative to the other nearby stars.

The photometry for these stars is shown in Figure \ref{fig:refs}.
None of these stars exhibit a large offset in either direction around 1100 days 
into the mission, at the time of the
rapid change in flux of \thisstar.
We can fit a long-term trend to each of these stars to measure the photometric long-term
variability
as recorded by \kep.
In practice, we fit five parameters: a slope, a zeropoint, and three relative offsets 
between observing seasons as the star falls on different detectors, maximizing
the likelihood of this model to the data. 
We also inspect each light curve by eye to look for significant deviations from
non-linearity similar to that detected for \thisstar.

For none of these stars do we recover a slope as extreme as $3.41 \pm 0.41$ ppt yr$^{-1}$.
Similarly, we do not detect any 2.5\% decreases in flux between days 1100 and 1250 (or at any other time), suggesting that
this effect observed in the light curve of \thisstar\ is not due to a spurious reference star, nor is there a systematic effect in all nearby stars.
We do note that we recover the same changes in photometric precision as observed for \thisstar\ as the stars move from
one detector to the next while the spacecraft rolls during its orbit.

We note that for some of the targets included in Figure \ref{fig:refs}, the
scatter between data points is larger than the photometric error bars. 
This is particularly true for KIC 8462696 and KIC 8462836. 
In these cases, the scatter is likely due to astrophysical variability rather
than an underestimation of our uncertainties.
For the former, long cadence photometry exists from the primary mission.
This star exhibits a clear photometric modulation due to starspots.
Here, spots modulate the flux of the star by 3\% with a rotation period of 16 days.
Because the rotation period is close to half the period of FFI observations, the
phase of the star's rotation when the telescope collects an FFI observation varies slowly,
leading to features such as the apparent trend in the FFI flux between day 400 and 800.
We note that this time period matches well with the peak of starspot activity for this star.
For \thisstar\ we know from the \kep\ data that the amplitude of photometric 
modulation is considerably smaller than the photometric uncertainties in the calibrated 
FFI data, so such a false positive trend cannot occur.

For KIC 8462836, we do not have \kep\ long cadence data. 
Photometry of the target is consistent with that expected of a mid-M dwarf.
For mid-M dwarfs, it is not uncommon to observe photometric variability caused
by starspots at the 3-5\% level \citep{Basri11}, so again, the excess scatter is likely the result of
starspot induced variability that should not be present for \thisstar.

\subsection{Other Nearby Stars}
\label{sec:nearbystars}
We can extend our analysis to search for variations in not only the most nearby stars used
as reference stars, but to all stars on the same detector.
We select all stars with $K_p$ within 0.5 mag of \thisstar\ and located in the same 
``skygroup,'' meaning they all fall on the same channel at the same time.
We remove stars that would not be expected to provide reliable tests in a search for
systematic trends in the \kep\ data, including known variable stars, eclipsing binaries,
and those with starspot-induced variability at the level of 1\%\ or more through a visual
inspection of the long cadence light curves.
These cuts leave 193 viable stars against which to compare.

\begin{figure}[htbp!]
\centerline{\includegraphics[width=0.45\textwidth]{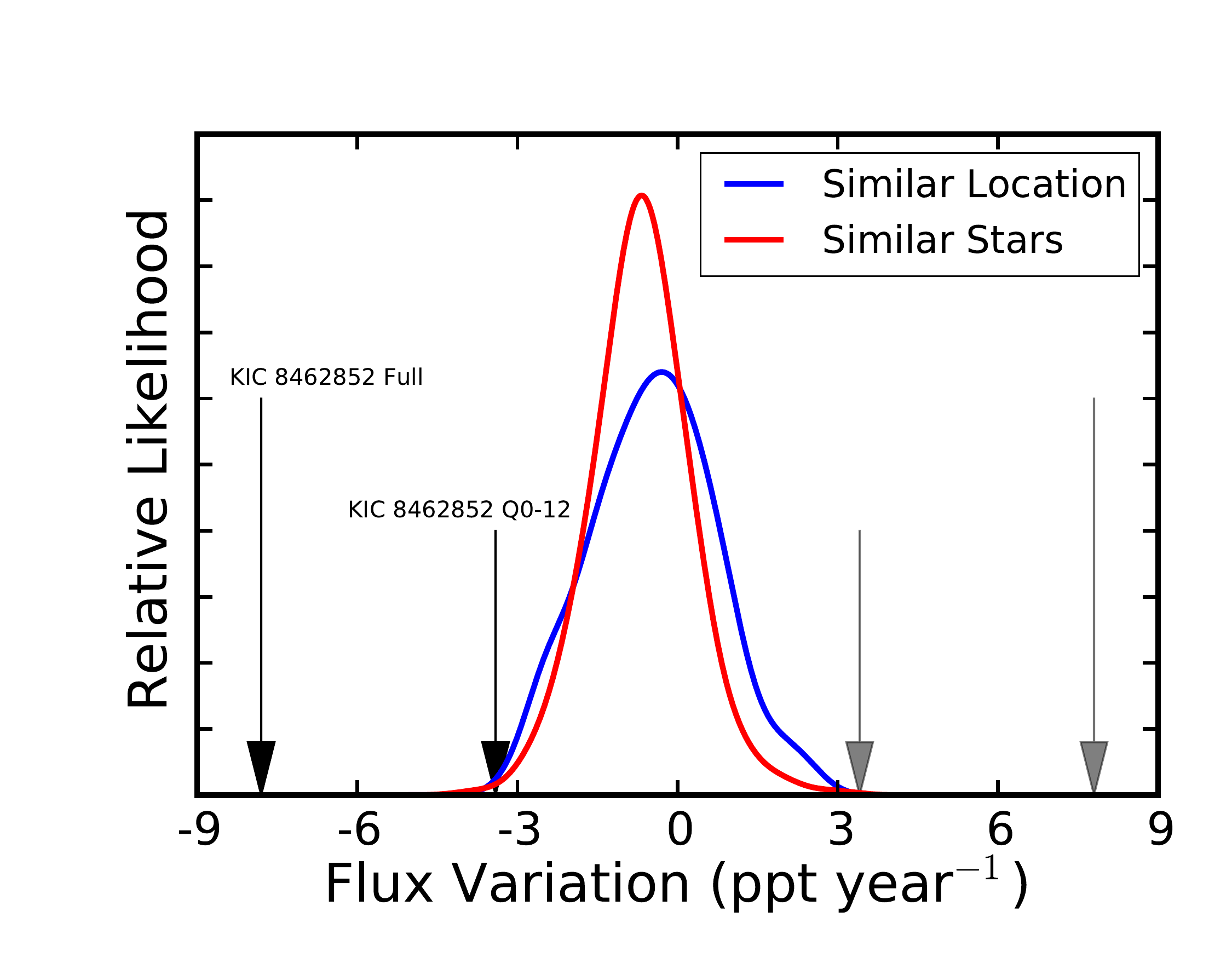}}
\caption{Kernel density estimator of the distribution of measured linear changes of the brightness of other stars observed in the \kep\ FFIs.
The red curve represents an analysis of other stars with similar properties to \thisstar;
the blue represents stars of similar magnitude on the same channel of the detector.
The bandwidth chosen in the creation of the KDE is equivalent to the uncertainty in the 
linear fit to the photometry of \thisstar\ over the first three years of the mission.
We find 0.6\% (0.3\%) of the mass of the KDE representing stars with similar properties
(similar locations)
is located at variations larger than those observed in the first three years of the light
curve of \thisstar\ (represented with a black arrow). 
No stars in the comparison sample produce slopes as large as would be measured by a simple 
linear fit to the full FFI light curve for our target (represented with a larger black arrow),
nor do we observe a rapid 2.5\% decrease in flux for any of these stars.
}
\label{fig:kde}
\end{figure}

We find, of these 193 stars, for only one does our automated pipeline measure a long-term trend: KIC 8395126
appears to decrease in brightness over the \kep\ mission, with a maximum likelihood measurement of the slope of 3.9 ppt yr$^{-1}$.
However, on closer inspection of the data for this star, we find that its
center of light is approximately one pixel from the position listed in the Kepler 
Target Database available on MAST. 
Reanalyzing this target after accounting for this one pixel offset causes the 
decay rate to decrease to 1.7 ppt yr$^{-1}$, half that of \thisstar.
We note that by repeating our analysis for \thisstar\ while moving the centroid in
any direction does not change the light curve beyond the level of the quoted photometric
uncertainties.
The measurements of the maximum likelihood slopes for the full sample of 193 nearby stars, again fitting for a slope and four offset
terms, are distributed with a mean of $-0.16$ ppt yr$^{-1}$ and a standard deviation
of 1.23 ppt yr$^{-1}$.
We plot a kernel density estimator (KDE) of the distribution of fitted slopes in Figure 
\ref{fig:kde}. For all cases, we apply a Gaussian kernel with a bandwidth of 0.41 
ppt yr$^{-1}$ ---  equal to the uncertainty in the measured Q0-Q12 slope for \thisstar\
and a reasonable estimate of the typical measurement uncertainty.

We find that none of the comparison stars and only 0.3\% of the mass of the KDE have a measured variation equal to
or larger than the 
maximum likelihood value of the slope of $3.41$ ppt yr$^{-1}$ recorded
over the first three years of the \kep\ mission for \thisstar. 
Even more significantly,
there is no mass at 7.8 ppt yr$^{-1}$, the value recovered if one were to blindly
fit a linear model to the distinctly non-linear light curve for \thisstar. 
In no cases do we observe a 2.5\% dip over 6 months as observed for \thisstar,
suggesting that the observed behavior of our target star is unique compared to all the comparison stars on the same detector.

\subsection{Other F stars}
\label{sec:Fstars}
We can also compare the flux variations observed for \thisstar\ to FFI light curves for
stars of similar spectral types.
Here, we use the sample of KIC stars developed by \citet{Lund16} and considered in their
analysis of the DASCH plates.
In that work, the authors selected 559 stars listed in the updated Kepler Input Catalog
\citep{Huber14} with inferred stellar effective temperatures
within 100 K, radii within 5\%, and $\log g$ within 10\% of \thisstar.

From this sample, we remove all stars for which we do not expect to be able to acquire reliable
photometry or observe long-term photometric trends over four-year baselines.
We remove stars with brighter neighbors located within ten pixels of the target star,
so that their flux is likely to leak into our own aperture.
We also remove stars near the edge of the detector and stars that do not fall onto the
\kep\ detectors in at least three of the four observing seasons. 
We remove stars that saturate the detector, and also those that are intrinsically variable,
including contact binaries, stars that vary due to spots at larger than the 2\% level,
and pulsating variable stars.
We note that the two targets that \citet{Lund16} noted as long-term variable, 
KIC 3868420 and KIC 11802860, both are removed by this last requirement.
The former exhibits 15\% variability with a five hour period in the \kep\ light curve; 
\citet{Nemec13} suggest it may be a high-amplitude Delta Scuti.
The latter exhibits 40\% variability with a 16.5 hour period in its \kep\ light curve;
the star is also known as AW Dra and has long been known as an RR Lyrae
\citep{Castellani98}.

The long-term variability observed by \citeauthor{Lund16} may be the result of unfortunate timing of the
observations across different phases of the star's light curve, causing an apparent long-term
trend.
This is plausible as the claimed photometric trends have different signs in the
pre- and post-1970 subsets of the DASCH data.
The long-term variability could also be a case of rapid stellar evolution:
\citet{Stellingwerf13} suggest RR Lyrae stars may be undergoing rapid mass-loss, which may
lead to a dimming over a timescale of decades.

We note that the overall distribution of F star variability is well-approximated by a Gaussian
with mean $-0.065$ ppt yr$^{-1}$ and standard deviation 0.094 ppt yr$^{-1}$. 
Stellar evolution does not explain a long-term trend in the typical F star; detection of a
nonzero mean dimming is likely indicative of a systematic in the \kep\ data that is not 
understood.
We repeat our analysis on these stars.
Again, none of the target stars exhibit a 3\% change in flux over the
duration of the \kep\ mission, nor do any stars change in flux by 2.5\% over 
six months.
Of all stars analyzed in this work, the rapid dimming between quarters 12 and 15 appears to be unique to 
\thisstar.
In fact, only a single other star, KIC 5868753, has an observed change in brightness
larger than the maximum likelihood slope for \thisstar\ from the first three
years of the \kep\ mission.

\section{Discussion}
\label{sec:discuss}

The \kep\ FFI light curve for \thisstar\ features a long-term slow dimming over the first
three years of the mission.
Formally, the observed slope is statistically significant, but from our analysis of
nearby stars it is likely that systematic uncertainties dominate over the
statistical uncertainties. 
There are no stars that we detect that exhibit a 7.8 ppt yr$^{-1}$ dimming,
the level that a simple linear model would produce when fit to the full, nonlinear data set.
Additionally, there are no stars that exhibit a 2.5\% decrement in flux over approximately 
six months.

Even if we consider only the relatively modest dimming over the first three years
of the \kep\ mission, there is only a single star that exhibits a slope that is at least as large in magnitude as the
one observed here.

The observed behavior of the star is thus very likely astrophysical in nature, 
suggesting this star is indeed undergoing some process leading to a
decrease in its observed brightness over the span of the \kep\ mission.  In the remainder of this section we consider possible interpretations of this result.

\subsection{Comparison to DASCH Photometry}

As mentioned in Section \ref{sec:intro}, \citet{Schaefer16} used 99 years of photometry from the DASCH project to analyze the behavior of \thisstar, finding a decrease in brightness of 14\%\ from 1890 to 1989.  \citet{Hippke16} performed an independent analysis of the DASCH photometry, confirming that the photographic data yield a fainter magnitude for \thisstar\ in the late 20th century compared to the end of the 19th century.  However, \citeauthor{Hippke16} argue that the DASCH measurements from 1890 to 1952 are best described by a constant brightness and measurements from 1967 to 1989 are best described by a different (fainter) constant brightness, with systematic errors accounting for the offset. We note that such systematic offsets have not been detected in DASCH light curves by the DASCH team \citep{Laycock10,Tang13a,Tang13b}. 

The unfortunate gap of 15 years with very few observations in the middle of the century makes it difficult to distinguish between a real astrophysical variation and a systematic offset.  Which explanation to prefer then depends on one's assessment of whether a star steadily fading for a century or a change in the photometric calibration of the Harvard plates around 1962 is more likely (or less unlikely).  The fading that we detect from 2009 to 2013 with the \kep\ FFI images does not necessarily represent a confirmation that \thisstar\ also dimmed over the preceding 129~yr, but could make that interpretation of the DASCH data more plausible.

Both previous analyses considered only linear models for the flux of the star as a function of time, as the DASCH data are insufficient to constrain more complex models.
The \kep\ FFI data are plainly nonlinear, suggesting that the true light curve
during the 20th century is likely more complex than either simple model.
Additionally, a decrease of 3.5\% in the flux of \thisstar\ is four times what would be expected
over the four-year \kep\ mission from \citeauthor{Schaefer16}'s century-long light curve.
Conversely, if the brightness change over the duration of the DASCH survey were as
extreme as that observed during the \kep\ mission, we would expect the star to have
decreased in brightness by 60\% over the 20th century, which does not appear to be consistent with comparison of archival and modern \citep[e.g.,][]{Boyajian16} measurements.
As there is a gap of two decades between the end of the DASCH data and the start of
the \kep\ mission, a direct comparison between the two is difficult.
Still, it is clear that neither a linear decrease in flux nor a constant flux model provide a good description of our observations, so there is no reason to expect the behavior at earlier times to have followed a simpler form.

\subsection{Instrumental Effects}

The observed dimming does not seem to be associated with any instrumental effect.
Some known effects could cause a 2.5\% decrement in the observed flux from a star.
The most common such effect is a Sudden Pixel Sensitivity Dropout (SPSD), in which a pixel
abruptly decreases in sensitivity, leading to a decrease in the observed flux on that pixel.
The pixel generally recovers in sensitivity over a few hours, but does not return to the
same sensitivity level as before \citep{Smith12}.
An SPSD could not explain our observed light curve,
as the observed flux decrement is visible in all four detector orientations, so the same 
SPSD would have had to occur on all four apertures, at the same level, at approximately
the same time.

Similarly, we do not expect this effect to be the result of any calibration issues during the
processing of the FFIs.
The calibration pipeline accounts for the bias and dark current of the detector,
cosmic rays, smearing effects from reading out without a shutter, and distortions induced from the readout electronics \citep{Quintana10}.
Most of these effects are dependent on the location of the star, rather than the physical
properties of the star itself, so we would not expect them to induce a dimming observed
over all four modules.
The exception is the smear correction, but the smear is well-understood and can be robustly
estimated from the data. 
Significant errors in the smear correction would be easily observable in the FFI data itself
as ``trails'' in the FFI image near the star, which are not detected in these observations.
Therefore, we can rule out instrumental effects or calibration errors as a plausible source of the observed variability.

\subsection{Comparison with Long-Cadence Light Curve}

The long cadence light curves produced by the \kep\ mission are accurate over
timescales of hours to days. Over longer timescales, systematics overwhelm long-term astrophysical
photometric variations. This is largely beause of two effects. First, there is a lack of reference stars, as only
$\approx 2\%$ of all stars in the \kep\ field are targeted for photometry at 30-minute cadence. Therefore, nearby stars that should share systematics are often not observed at the
cadence that would be necessary for co-trending to identify shared systematics. 
Second, the apertures downloaded by \kep\ are typically smaller than the PSF of each
individual star (Figure 2), so that small changes in the pointing of the telescope can
cause significant flux variations, overwhelming any signal from the star itself.
With the FFI data in hand, we create our own long-cadence light curve from the pixel-level data.

As the FFI data provide absolute calibration of the \kep\ data once per month, the FFI data 
can be combined with the flattened \kep\ long cadence light curve to 
show one plausible iteration of the true astrophysical long-cadence light curve of \thisstar, with instrumental effects removed. 
To create this light curve, we fit a cubic spline to the FFI data using the spline tool in
\texttt{scipy.optimize}. 
We use the measured photometric precision for each FFI as the input weights, applying a smoothing factor of 50. 
The resultant smooth light curve is then multiplied by the flattened light curve of
Section 3.3 to create a realistic long-cadence light curve that captures
both the short-term variability and the long-term dimming observed during the
\kep\ mission.
These data are plotted in Figure \ref{fig:photometry} and included in the same ``Data Behind the Figure'' as both the intermediate, flattened light curve described in the previous paragraph and the smooth spline fit to the FFI data.
As the data are not calibrated between the FFIs, and each FFI observation has an associated
photometric uncertainty,
this long-cadence light curve represents one possible realization of the photometric variability of \thisstar\ and should only be considered in that context.

\subsection{Background Contamination}
Approximately 25 arcsec from \thisstar\ is KIC 8462860, a star 3.65 magnitudes fainter
on the edge of our photometric aperture.
As approximately half of the star falls in our simple aperture, we would expect it to
contribute approximately 1.8\% of the total flux in our aperture.
This star cannot explain the long-term trend over the first three years of the mission
unless it were to decrease in flux by 50\%; it cannot explain the rapid dimming event
unless it disappeared entirely (which is obviously not the case).

KIC 8462860 could explain the observed rapid dimming if the star had a large proper 
motion and moved out of the aperture during Quarters 13 and 14.
However, in the PPMXL catalog \citep{ppmxl} its proper motion is $\mu_{\rm RA} = 2.4 \pm 4.1$~mas~yr$^{-1}$, $\mu_{\rm Dec} = 0.3 \pm 4.1$~mas~yr$^{-1}$.
The first \textit{Gaia} data release does not include a proper
motion measurement of KIC 8462860. By combining the \textit{Gaia}
measurement of the position of the star \citep{Lindegren16} with
the position recorded in the 2MASS point source catalog 
\citep{Cutri03}, we find the proper motion must be less than
$\approx 6$~mas~yr$^{-1}$, in line with the PPMXL result.
The star is listed in the KIC as having a temperature of 5464 K, and its magnitude and
colors ($m_{r}$ = 15.6, $r-J$ = 1.4, $J-K$ = 0.5) are broadly consistent with a late G or early K dwarf at several hundred parsec or an evolved G/K star at larger distances.
The photometry is not consistent with what would be expected of a nearby star.

If the star did have a large proper motion, we would see it move by several pixels
over the course of the mission, which visual inspection of the FFI data does not show.
Moreover, to move far enough so that the entire core of the PSF moved across the edge
of our aperture over a span of approximately six months, the proper motion would need to
be approaching 10 arcsec yr$^{-1}$. 
In this case, we would expect the star to traverse the entire aperture over the four years
of FFI data, which we do not observe.

Perhaps most significantly, the same light curve as shown in Figure \ref{fig:photometry}
is recovered when we mask the pixels corresponding to KIC 8462860, or if we modify the size of
our aperture to either move this other star fully inside or outside of the aperture rather
than on the edge.
Therefore, we can exclude the possibility that the observed variations in the light curve
are caused by the presence of KIC 8462860.
Moreover, we do not detect the presence of any centroid shifts correlated with the measured
flux in the light curve of this star, suggesting this result is not the effect of the neighboring
star or changes in the underlying flat field.
We can rule out background contamination as explaining our light curve
with high confidence.

\subsection{Transiting Material}
\label{sec:transit}

We note that the shape of the light curve after Quarter 12 appears broadly similar to that of a
transit event.
Over six months, the rapid dimming could represent the ingress of material blocking the stellar
disk, leading to a decrease in observed flux. 
The primary mission then ends before third contact is observed.
Clouds of transiting material due to disintegrating bodies have been
observed in \kep\ data, but only with very short orbital periods
\citep{Rappaport14, SanchisOjeda15}.
In the case of \thisstar, there are problems with this model.
First, it does not explain the observed long-term dimming over the first three years of the
\kep\ mission, nor does it explain the long-term trend observed over the past 100 years
through the DASCH data.
A single transit model also does not explain the phenomena observed by \citet{Boyajian16}, especially those
dips before the time of ingress.  Nevertheless, we investigate whether it is a possible explanation for the rapid dimming.

A transit event with a timescale similar to that observed here is not unprecedented.
$\epsilon$~Aurigae is transited every 27 years, with a transit lasting approximately 
1.5 years \citep{Kopal54, Huang65}.
Recently, \citet{Rodriguez16} observed transits of TYC~2505-672-1 with a period of
69.1 years and a duration of 3.45 years. 
In both cases, the primary is orbited by a hot source that hosts an extended disk of
circumstellar material.
\citet{Kloppenborg10} confirmed this model with closure-phase interferometric imaging,
directly imaging the disk occulting $\epsilon$~Aurigae.

For both $\epsilon$~Aurigae and TYC~2505-672-1, the disk is optically thick and circumstellar around a binary companion.  The sub-millimeter observations of \citet{Thompson16} place a limit of 7.7 \mearth\ of material within 200 AU of \thisstar, ruling out a direct analogue of these other systems.

First, we consider that this signal represents the transit of a solid 
body across the stellar disk of \thisstar.
In this scenario the observed timescales of the event constrain the size and distance of the eclipsing object.  For an optically thick transiting object, the 2.5\%\ transit depth indicates a minimum radius of 0.15~R$_{*}$ (\citealt{Boyajian16} estimate a radius of 1.58~R$_{\odot}$ for \thisstar).  If the transiting body is in a Keplerian orbit, the extremely slow ingress time and long transit duration place it at the implausibly large distance of $\sim10$~pc, with a transit probability of $\sim 10^{-9}$.  Note that while a six-month transit ingress and transit duration of more than six months are similar to $\epsilon$~Aurigae and TYC~2505-672-1, this case is distinct from those because of the much smaller radius of \thisstar.  The transit of a main sequence star is necessarily much shorter than that of a supergiant with $R \gtrsim 50$~R$_{\odot}$.  Even if every star had a companion at 10 pc, then given $10^4$ \kep\ missions we would expect to observe one such transit event; we can confidently disfavor this hypothesis.

If the transiting material is instead an optically thin cloud, then its size could be comparable to or larger than that of \thisstar.  The optical depth of such a cloud would be $\tau \approx 0.025$.  Because the orbiting body considered in the previous paragraph was already a significant fraction of the stellar radius, though, a much bigger cloud does not qualitatively change the result: in order for the transit ingress to last $\sim180$~d, a cloud in a Keplerian orbit would need to be located $\gtrsim1$~pc away from the star.  We thus conclude that a simple transit of any kind of orbiting object is not a reasonable explanation for the rapid fading that begins in Quarter 12.  However, this conclusion relies on the assumption that the duration of the transit ingress is set by the orbital velocity of the transiting body.  

More complex scenarios in which the ingress timescale reflects the spreading of debris along its orbit after a recent collision or the precession of an occulting disk into our line of sight could perhaps explain the appearance of a transit.
For example, a cloud that slowly increases in density would manifest itself in the
light curve as inducing a change in flux similar to that observed in Quarter 12.
However, to produce the apparent flat bottom of the supposed transit event, such a cloud
would then need to be extended over a fraction of its Keplerian orbit and 
would need to maintain an approximately constant density through its entire length as
it passes in front of this star.

As illustrated above,
it is very difficult to come up with a physical model that can even qualitatively explain all of the major features of \thisstar's photometric behavior simultaneously.  Of the ideas that have been proposed so far, we suggest that the most promising explanation involves a recent collision between large bodies (planetesimals or comets) in the \thisstar\ system,
leading to a spreading of debris as in the previous paragraph. 
In this picture, a recent collision could create a cloud of circumstellar material and push
a family of objects into a highly eccentric orbit, analogous to the period of late heavy
bombardment observed in our own solar system.  
However, this idea does not naturally account for the steady decline in the flux of \thisstar\ in the years preceding the more rapid dimming and the concentrated sequence of dips.
The data presented in this paper cannot fully exclude any of these models, but we note that the circumstellar dust and debris produced by such an event is unlikely to maintain this arrangement for long timescales.
The sub-mm limits of \citet{Thompson16} and continued photometric monitoring will significantly constrain
future models that attempt to invoke circumstellar dust or transiting models to explain this light curve.
Transiting material remains as a plausible explanation for
the \thisstar\ light curve, but requires particular, \textit{a priori} unlikely density profiles for the circumstellar
material in order to match the data.

\subsection{A Polar Spot}
Under certain conditions, a long-lasting spot growing at polar latitudes
on the surface of \thisstar\ could possibly reproduce the long-term light curve
observed here.
While polar spots have not previously been detected on an F3V dwarf star,
they have been observed on the surface of an F9V
star, albeit one in a tight (1.15 day) binary
with another stellar companion \citep{Strassmeier03}.
Through interferometric aperture synthesis imaging, polar spots on more evolved
stars have been seen to evolve on similar timescales to the flux variability observed here \citep{Roettenbacher16}. 

For a polar spot to create the observed decrease in flux in the light
curve, the projection of the polar region onto the observer's line of sight would need to be large enough to allow for a large starspot to be observed.
At high (edge-on) inclinations, a polar spot would be foreshortened, diminishing
its effect on the light curve. 
Moreover, limb darkening would decrease the overall contribution of the polar regions
on the light curve.
At lower (more face-on) inclinations, the pole would be always visible and a polar spot
would have a larger effect on the overall light curve.
\citet{Boyajian16} measure a rotation period and $v \sin i$ of the star
and combine these with an estimate of the radius to infer an inclination
of $68 \pm 29$ degrees at 68\% confidence, leaving open the possibility
that the pole can indeed be observed at all times well away from the edge of the
stellar disk.

If a spot were growing near the polar latitudes of the surface of the
star between Quarters 13 and 15, we might expect a corresponding
increase in overall magnetic activity, leading to an increase in 
starspots at other latitudes over this time period.
This indeed appears to be the case. 
\citet{Boyajian16} measure a periodic signal with a period of 0.88 days, as well as additional signals at 0.90 and 0.96 days, all of which change in intensity during the \kep\ mission. 
The authors suggest that this signal may be induced by starspots.
Spots at different latitudes evolving and rotating differentially would produce a signal like this one.
The strength of the signal they observe grows between days 1100 and 1300 of the mission, suggesting either an increase in the number of spots or an increase in their coherence. 
Interestingly, the growth of the signal corresponds to
the time of rapid decrease in total flux recorded in the FFI images.

The inclination of \thisstar\ and the coincidence between the spot activity and the rapid dimming do not present a confirmation of a polar spot.
A long-lasting polar spot itself would be remarkable given the F3V spectral type of the star, but with the current
data the hypothesis cannot be excluded.
Broadly, the FFI observations are consistent with the growth of a polar spot that grows at a 
similar time to the growth of spots at lower latitudes. Spots, however, 
can not explain the short-term dips originally observed by \citet{Boyajian16}. More observations
are needed to separate the spot hypothesis from other possible explanations of the observed light curve.
As the polar spot hypothesis cannot account for the short-term dips of \citet{Boyajian16}, it does not seem to be a particularly likely scenario, but
cannot be ruled out given the available data.

\subsection{Additional Observations}

In this work, we do not present a model that can explain the entire suite of observations
of \thisstar. 
There are now three distinct photometric variations observed: rapid, irregular decreases
of $10\%$ or more in flux lasting for a few days, a $2.5\%$ decrease in flux lasting
for at least one year, and a likely long-term dimming perhaps spanning more than a
century. 
Additional observations would be helpful in order to better understand physical 
phenomena that could cause any or all of these events.

Multi-color photometry is essential to help characterize this star. If any of these events are caused by solid bodies, we would expect the photometric variations to be largely
achromatic. 
However, if they are caused by a cloud of dust and gas, we would expect the cloud
to redden the star. 
Similarly, spots with a lower effective temperature than the rest
of the stellar surface would cause an apparent change in the color of the
star as they cover more of the stellar disc.
A transiting cloud would be expected to induce dips that are largely periodic, while changes in the spot patterns would not necessarily be periodic.
High precision observations of the color of \thisstar\ over time could explain the nature of each of these events, supporting or ruling out the transiting cloud or spot hypotheses.
Similarly, additional IR and sub-mm observations of the system could be used to place
tighter upper limits on the amount of circumstellar material surrounding the star.

Additional radial velocity (RV) monitoring to search for companions in few-AU orbits around the star,
especially those that could hold together a disk of circumstellar material into a 
gravitationally bound system, would also be useful.
\citet{Boyajian16} obtained four high resolution spectra over the course of approximately 500 days.
The measured RV of \thisstar\ in these observations has a scatter of $0.3$ km s$^{-1}$ and
each observation has a precision of $0.4$ km s$^{-1}$.
While the data are consistent with no RV variations, continued monitoring could detect
the presence of massive companions on wider orbits.
High resolution spectra could also be used to probe any evolution in the
magnetic activity of the star correlated with the growth or decay
of spots.

\section{Conclusions}
\label{sec:conclusions}

Recently, multiple analyses of DASCH photometry have produced conflicting results
about the possible detection of a long-term dimming of \thisstar\ by $0.165 \pm 0.013$ magnitudes over the 20th century, or $0.152 \pm 0.012\%$~yr$^{-1}$.
The dimming of a star at that rate should be detectable in \kep\ data.
Here, we analyze monthly \kep\ FFI images to search for similar dimming in the FFI light curve
of \thisstar. 

We perform aperture photometry on \thisstar\ and seven nearby comparison stars.
We observe that during the first three years of the \kep\ mission, the star dimmed at a rate
of $0.341 \pm 0.041$\% yr$^{-1}$.
Over the following 6 months, the star decreased in brightness by 2.5\%, then remained at that level
for the duration of the primary \kep\ mission.
This result is not sensitive to the size of the chosen aperture or the particular choice
of reference stars.

We then compare this result to a similar analysis of other stars of similar brightness on the
same detector, as well as stars with similar stellar properties, as listed in the KIC, in the
\kep\ field.
We find that 0.3\% of stars on the same detector and 0.6\% of stars with similar stellar
properties exhibit a long-term trend consistent with that observed for \thisstar\ during the first
three years of the \kep\ mission. However, in no cases do we observe a flux decrement as
extreme as the 2.5\% dip observed in Quarters 12-14 of the mission.  The total brightness change of \thisstar\ is also larger than that of any other star we have identified in the \kep\ images.

Broadly speaking, the morphology of the light curve is generally consistent with the transit
of a cloud of optically thick material orbiting the star.
Such a dust cloud could be small enough to evade submillimeter detection in the analysis
of \citet{Thompson16}, who place a limit of 7.7 \mearth\ of material orbiting \thisstar.
The breakup of a small body or a recent collision that could produce a cloud of material could also plausibly produce a family of comets that transit the
host star together as one group \citep{Bodman16}, explaining the light curve observed by \citet{Boyajian16}.
However, in order to match the observed time of ingress and transit duration, some fine-tuning is required.
To explain the transit ingress timescale, the cloud would need to be at impossibly large
distances from the star or be slowly increasing in surface density. 
The flat bottom of the transit would then suggest a rapid transition into a region of uniform density in the cloud, which then continues to transit the star for at least
the next year of the \kep\ mission.
Moreover, such a model does not naturally account for the long-term dimming in the light curve 
observed in both DASCH and the \kep\ FFI data, suggesting that this idea is, at best, 
incomplete.  

There is no known or proposed stellar phenomenon that can fully explain 
all aspects of the observed light curve. Non-stellar
explanations, such as circumstellar material, offer the best opportunity to 
provide an explanation for the observed light curve, but simple models are unable
to match the light curve as observed over the last century.
We strongly encourage further refinements, alternative hypotheses, and new data in order to explain the full suite of observations of this very mysterious object.

{\it Facility:} \facility{Kepler}

\acknowledgements

We thank Jason Dittmann and Jieun Choi (Harvard) for conversations about data analysis and figure design that improved the quality of this paper. We thank John Brewer (Yale/Columbia) for finding a typo in Equation 1
in an earlier version of this paper. We also thank 
Rachael Roettenbacher (Stockholm University) and John Johnson (Harvard) for discussions of starspots, Ryan Foley (UC Santa Cruz) for initial conversations about photometry and the DASCH results, George Preston (Carnegie) for insights into variable stars, and Eugene Chiang (UC Berkeley) and Jason Wright (Penn State) for ideas about other potentially similar systems.

B.T.M. is supported by the National Science Foundation Graduate Research Fellowship under Grant No. DGE-1144469.

Funding for \kep, the tenth Discovery mission, was
provided by NASA's Science Mission Directorate. 
We are grateful to the entire \kep\ team, past and present.
Their tireless efforts were essential to the tremendous success of the mission and the
successes of \textit{K2}, present and future.

All of the data presented in this paper were obtained from the Mikulski Archive for Space Telescopes (MAST). STScI is operated by the Association of Universities for Research in Astronomy, Inc., under NASA contract NAS5-26555. Support for MAST for non-HST data is provided by the NASA Office of Space Science via grant NNX09AF08G and by other grants and contracts.

This research has made use of the SIMBAD database,
operated at CDS, Strasbourg, France and NASA's Astrophysics Data System.


\begin{thebibliography}{}
\expandafter\ifx\csname natexlab\endcsname\relax\def\natexlab#1{#1}\fi

\bibitem[{{Abeysekara} {et~al.}(2016){Abeysekara}, {Archambault}, {Archer},
  {Benbow}, {Bird}, {Buchovecky}, {Buckley}, {Byrum}, {Cardenzana}, {Cerruti},
  {Chen}, {Christiansen}, {Ciupik}, {Cui}, {Dickinson}, {Eisch}, {Errando},
  {Falcone}, {Fegan}, {Feng}, {Finley}, {Fleischhack}, {Fortin}, {Fortson},
  {Furniss}, {Gillanders}, {Griffin}, {Grube}, {Gyuk}, {H{\"u}tten},
  {H{\aa}kansson}, {Hanna}, {Holder}, {Humensky}, {Johnson}, {Kaaret}, {Kar},
  {Kelley-Hoskins}, {Kertzman}, {Kieda}, {Krause}, {Krennrich}, {Kumar},
  {Lang}, {Lin}, {Maier}, {McArthur}, {McCann}, {Meagher}, {Moriarty},
  {Mukherjee}, {Nieto}, {O'Brien}, {O'Faol{\'a}in de Bhr{\'o}ithe}, {Ong},
  {Otte}, {Park}, {Perkins}, {Petrashyk}, {Pohl}, {Popkow}, {Pueschel},
  {Quinn}, {Ragan}, {Ratliff}, {Reynolds}, {Richards}, {Roache}, {Santander},
  {Sembroski}, {Shahinyan}, {Staszak}, {Telezhinsky}, {Tucci}, {Tyler},
  {Vincent}, {Wakely}, {Weiner}, {Weinstein}, {Williams}, \&
  {Zitzer}}]{Abeysekara16}
{Abeysekara}, A.~U., {Archambault}, S., {Archer}, A., {et~al.} 2016, \apjl,
  818, L33

\bibitem[{{Barnes} {et~al.}(2016){Barnes}, {Weingrill}, {Fritzewski},
  {Strassmeier}, \& {Platais}}]{Barnes16}
{Barnes}, S.~A., {Weingrill}, J., {Fritzewski}, D., {Strassmeier}, K.~G., \&
  {Platais}, I. 2016, \apj, 823, 16

\bibitem[{{Basri} {et~al.}(2011){Basri}, {Walkowicz}, {Batalha}, {Gilliland},
  {Jenkins}, {Borucki}, {Koch}, {Caldwell}, {Dupree}, {Latham}, {Marcy},
  {Meibom}, \& {Brown}}]{Basri11}
{Basri}, G., {Walkowicz}, L.~M., {Batalha}, N., {et~al.} 2011, \aj, 141, 20

\bibitem[{{Bastien} {et~al.}(2013){Bastien}, {Stassun}, {Basri}, \&
  {Pepper}}]{Bastien13}
{Bastien}, F.~A., {Stassun}, K.~G., {Basri}, G., \& {Pepper}, J. 2013, \nat,
  500, 427

\bibitem[{{Bedding} {et~al.}(2011){Bedding}, {Mosser}, {Huber},
  {Montalb{\'a}n}, {Beck}, {Christensen-Dalsgaard}, {Elsworth},
  {Garc{\'{\i}}a}, {Miglio}, {Stello}, {White}, {De Ridder}, {Hekker}, {Aerts},
  {Barban}, {Belkacem}, {Broomhall}, {Brown}, {Buzasi}, {Carrier}, {Chaplin},
  {di Mauro}, {Dupret}, {Frandsen}, {Gilliland}, {Goupil}, {Jenkins},
  {Kallinger}, {Kawaler}, {Kjeldsen}, {Mathur}, {Noels}, {Silva Aguirre}, \&
  {Ventura}}]{Bedding11}
{Bedding}, T.~R., {Mosser}, B., {Huber}, D., {et~al.} 2011, \nat, 471, 608

\bibitem[{{Bodman} \& {Quillen}(2016)}]{Bodman16}
{Bodman}, E.~H.~L., \& {Quillen}, A. 2016, \apjl, 819, L34

\bibitem[{{Borucki} {et~al.}(2010){Borucki}, {Koch}, {Basri}, {Batalha},
  {Brown}, {Caldwell}, {Caldwell}, {Christensen-Dalsgaard}, {Cochran},
  {DeVore}, {Dunham}, {Dupree}, {Gautier}, {Geary}, {Gilliland}, {Gould},
  {Howell}, {Jenkins}, {Kondo}, {Latham}, {Marcy}, {Meibom}, {Kjeldsen},
  {Lissauer}, {Monet}, {Morrison}, {Sasselov}, {Tarter}, {Boss}, {Brownlee},
  {Owen}, {Buzasi}, {Charbonneau}, {Doyle}, {Fortney}, {Ford}, {Holman},
  {Seager}, {Steffen}, {Welsh}, {Rowe}, {Anderson}, {Buchhave}, {Ciardi},
  {Walkowicz}, {Sherry}, {Horch}, {Isaacson}, {Everett}, {Fischer}, {Torres},
  {Johnson}, {Endl}, {MacQueen}, {Bryson}, {Dotson}, {Haas}, {Kolodziejczak},
  {Van Cleve}, {Chandrasekaran}, {Twicken}, {Quintana}, {Clarke}, {Allen},
  {Li}, {Wu}, {Tenenbaum}, {Verner}, {Bruhweiler}, {Barnes}, \&
  {Prsa}}]{Borucki10}
{Borucki}, W.~J., {Koch}, D., {Basri}, G., {et~al.} 2010, Science, 327, 977

\bibitem[{{Boyajian} {et~al.}(2012){Boyajian}, {von Braun}, {van Belle},
  {McAlister}, {ten Brummelaar}, {Kane}, {Muirhead}, {Jones}, {White},
  {Schaefer}, {Ciardi}, {Henry}, {L{\'o}pez-Morales}, {Ridgway}, {Gies}, {Jao},
  {Rojas-Ayala}, {Parks}, {Sturmann}, {Sturmann}, {Turner}, {Farrington},
  {Goldfinger}, \& {Berger}}]{Boyajian12}
{Boyajian}, T.~S., {von Braun}, K., {van Belle}, G., {et~al.} 2012, \apj, 757,
  112

\bibitem[{{Boyajian} {et~al.}(2016){Boyajian}, {LaCourse}, {Rappaport},
  {Fabrycky}, {Fischer}, {Gandolfi}, {Kennedy}, {Korhonen}, {Liu}, {Moor},
  {Olah}, {Vida}, {Wyatt}, {Best}, {Brewer}, {Ciesla}, {Cs{\'a}k}, {Deeg},
  {Dupuy}, {Handler}, {Heng}, {Howell}, {Ishikawa}, {Kov{\'a}cs}, {Kozakis},
  {Kriskovics}, {Lehtinen}, {Lintott}, {Lynn}, {Nespral}, {Nikbakhsh},
  {Schawinski}, {Schmitt}, {Smith}, {Szabo}, {Szabo}, {Viuho}, {Wang},
  {Weiksnar}, {Bosch}, {Connors}, {Goodman}, {Green}, {Hoekstra}, {Jebson},
  {Jek}, {Omohundro}, {Schwengeler}, \& {Szewczyk}}]{Boyajian16}
{Boyajian}, T.~S., {LaCourse}, D.~M., {Rappaport}, S.~A., {et~al.} 2016,
  \mnras, 457, 3988

\bibitem[{{Bryson} {et~al.}(2010){Bryson}, {Tenenbaum}, {Jenkins},
  {Chandrasekaran}, {Klaus}, {Caldwell}, {Gilliland}, {Haas}, {Dotson}, {Koch},
  \& {Borucki}}]{Bryson10}
{Bryson}, S.~T., {Tenenbaum}, P., {Jenkins}, J.~M., {et~al.} 2010, \apjl, 713,
  L97

\bibitem[{{Butler} {et~al.}(1996){Butler}, {Marcy}, {Williams}, {McCarthy},
  {Dosanjh}, \& {Vogt}}]{Butler96b}
{Butler}, R.~P., {Marcy}, G.~W., {Williams}, E., {et~al.} 1996, \pasp, 108, 500

\bibitem[{{Caldwell} {et~al.}(2010){Caldwell}, {Kolodziejczak}, {Van Cleve},
  {Jenkins}, {Gazis}, {Argabright}, {Bachtell}, {Dunham}, {Geary}, {Gilliland},
  {Chandrasekaran}, {Li}, {Tenenbaum}, {Wu}, {Borucki}, {Bryson}, {Dotson},
  {Haas}, \& {Koch}}]{Caldwell10}
{Caldwell}, D.~A., {Kolodziejczak}, J.~J., {Van Cleve}, J.~E., {et~al.} 2010,
  \apjl, 713, L92

\bibitem[{{Carter} {et~al.}(2012){Carter}, {Agol}, {Chaplin}, {Basu},
  {Bedding}, {Buchhave}, {Christensen-Dalsgaard}, {Deck}, {Elsworth},
  {Fabrycky}, {Ford}, {Fortney}, {Hale}, {Handberg}, {Hekker}, {Holman},
  {Huber}, {Karoff}, {Kawaler}, {Kjeldsen}, {Lissauer}, {Lopez}, {Lund},
  {Lundkvist}, {Metcalfe}, {Miglio}, {Rogers}, {Stello}, {Borucki}, {Bryson},
  {Christiansen}, {Cochran}, {Geary}, {Gilliland}, {Haas}, {Hall}, {Howard},
  {Jenkins}, {Klaus}, {Koch}, {Latham}, {MacQueen}, {Sasselov}, {Steffen},
  {Twicken}, \& {Winn}}]{Carter12}
{Carter}, J.~A., {Agol}, E., {Chaplin}, W.~J., {et~al.} 2012, Science, 337, 556

\bibitem[{{Castellani} {et~al.}(1998){Castellani}, {di Paolantonio},
  {Piersimoni}, \& {Ripepi}}]{Castellani98}
{Castellani}, V., {di Paolantonio}, A., {Piersimoni}, A.~M., \& {Ripepi}, V.
  1998, \aap, 333, 918

\bibitem[{{Chaplin} {et~al.}(2014){Chaplin}, {Basu}, {Huber}, {Serenelli},
  {Casagrande}, {Silva Aguirre}, {Ball}, {Creevey}, {Gizon}, {Handberg},
  {Karoff}, {Lutz}, {Marques}, {Miglio}, {Stello}, {Suran}, {Pricopi},
  {Metcalfe}, {Monteiro}, {Molenda-{\.Z}akowicz}, {Appourchaux},
  {Christensen-Dalsgaard}, {Elsworth}, {Garc{\'{\i}}a}, {Houdek}, {Kjeldsen},
  {Bonanno}, {Campante}, {Corsaro}, {Gaulme}, {Hekker}, {Mathur}, {Mosser},
  {R{\'e}gulo}, \& {Salabert}}]{Chaplin14}
{Chaplin}, W.~J., {Basu}, S., {Huber}, D., {et~al.} 2014, \apjs, 210, 1

\bibitem[{{Cutri} {et~al.}(2003){Cutri}, {Skrutskie}, {van Dyk}, {Beichman},
  {Carpenter}, {Chester}, {Cambresy}, {Evans}, {Fowler}, {Gizis}, {Howard},
  {Huchra}, {Jarrett}, {Kopan}, {Kirkpatrick}, {Light}, {Marsh}, {McCallon},
  {Schneider}, {Stiening}, {Sykes}, {Weinberg}, {Wheaton}, {Wheelock}, \&
  {Zacarias}}]{Cutri03}
{Cutri}, R.~M., {Skrutskie}, M.~F., {van Dyk}, S., {et~al.} 2003, {2MASS All
  Sky Catalog of point sources.}

\bibitem[{{Finkbeiner} {et~al.}(2016){Finkbeiner}, {Schlafly}, {Schlegel},
  {Padmanabhan}, {Juri{\'c}}, {Burgett}, {Chambers}, {Denneau}, {Draper},
  {Flewelling}, {Hodapp}, {Kaiser}, {Magnier}, {Metcalfe}, {Morgan}, {Price},
  {Stubbs}, \& {Tonry}}]{Finkbeiner16}
{Finkbeiner}, D.~P., {Schlafly}, E.~F., {Schlegel}, D.~J., {et~al.} 2016, \apj,
  822, 66

\bibitem[{{Garc{\'{\i}}a} {et~al.}(2011){Garc{\'{\i}}a}, {Hekker}, {Stello},
  {Guti{\'e}rrez-Soto}, {Handberg}, {Huber}, {Karoff}, {Uytterhoeven},
  {Appourchaux}, {Chaplin}, {Elsworth}, {Mathur}, {Ballot},
  {Christensen-Dalsgaard}, {Gilliland}, {Houdek}, {Jenkins}, {Kjeldsen},
  {McCauliff}, {Metcalfe}, {Middour}, {Molenda-Zakowicz}, {Monteiro}, {Smith},
  \& {Thompson}}]{Garcia11}
{Garc{\'{\i}}a}, R.~A., {Hekker}, S., {Stello}, D., {et~al.} 2011, \mnras, 414,
  L6

\bibitem[{{Gilliland} {et~al.}(2011){Gilliland}, {Chaplin}, {Dunham},
  {Argabright}, {Borucki}, {Basri}, {Bryson}, {Buzasi}, {Caldwell}, {Elsworth},
  {Jenkins}, {Koch}, {Kolodziejczak}, {Miglio}, {van Cleve}, {Walkowicz}, \&
  {Welsh}}]{Gilliland11}
{Gilliland}, R.~L., {Chaplin}, W.~J., {Dunham}, E.~W., {et~al.} 2011, \apjs,
  197, 6

\bibitem[{{Grindlay} {et~al.}(2009){Grindlay}, {Tang}, {Simcoe}, {Laycock},
  {Los}, {Mink}, {Doane}, \& {Champine}}]{grindlay09}
{Grindlay}, J., {Tang}, S., {Simcoe}, R., {et~al.} 2009, in Astronomical
  Society of the Pacific Conference Series, Vol. 410, Preserving Astronomy's
  Photographic Legacy: Current State and the Future of North American
  Astronomical Plates, ed. W.~{Osborn} \& L.~{Robbins}, 101

\bibitem[{{Haas} {et~al.}(2010){Haas}, {Batalha}, {Bryson}, {Caldwell},
  {Dotson}, {Hall}, {Jenkins}, {Klaus}, {Koch}, {Kolodziejczak}, {Middour},
  {Smith}, {Sobeck}, {Stober}, {Thompson}, \& {Van Cleve}}]{Haas10}
{Haas}, M.~R., {Batalha}, N.~M., {Bryson}, S.~T., {et~al.} 2010, \apjl, 713,
  L115

\bibitem[{{Harp} {et~al.}(2016){Harp}, {Richards}, {Shostak}, {Tarter},
  {Vakoch}, \& {Munson}}]{Harp15}
{Harp}, G.~R., {Richards}, J., {Shostak}, S., {et~al.} 2016, \apj, 825, 155

\bibitem[{{Hippke} {et~al.}(2016){Hippke}, {Angerhausen}, {Lund}, {Pepper}, \&
  {Stassun}}]{Hippke16}
{Hippke}, M., {Angerhausen}, D., {Lund}, M.~B., {Pepper}, J., \& {Stassun},
  K.~G. 2016, \apj, 825, 73

\bibitem[{{Huang}(1965)}]{Huang65}
{Huang}, S.-S. 1965, \apj, 141, 976

\bibitem[{{Huber} {et~al.}(2014){Huber}, {Silva Aguirre}, {Matthews},
  {Pinsonneault}, {Gaidos}, {Garc{\'{\i}}a}, {Hekker}, {Mathur}, {Mosser},
  {Torres}, {Bastien}, {Basu}, {Bedding}, {Chaplin}, {Demory}, {Fleming},
  {Guo}, {Mann}, {Rowe}, {Serenelli}, {Smith}, \& {Stello}}]{Huber14}
{Huber}, D., {Silva Aguirre}, V., {Matthews}, J.~M., {et~al.} 2014, \apjs, 211,
  2

\bibitem[{{Jenkins} {et~al.}(2010){Jenkins}, {Caldwell}, {Chandrasekaran},
  {Twicken}, {Bryson}, {Quintana}, {Clarke}, {Li}, {Allen}, {Tenenbaum}, {Wu},
  {Klaus}, {Middour}, {Cote}, {McCauliff}, {Girouard}, {Gunter}, {Wohler},
  {Sommers}, {Hall}, {Uddin}, {Wu}, {Bhavsar}, {Van Cleve}, {Pletcher},
  {Dotson}, {Haas}, {Gilliland}, {Koch}, \& {Borucki}}]{Jenkins10}
{Jenkins}, J.~M., {Caldwell}, D.~A., {Chandrasekaran}, H., {et~al.} 2010,
  \apjl, 713, L87

\bibitem[{{Kinemuchi}(2011)}]{Kinemuchi11}
{Kinemuchi}, K. 2011, in RR Lyrae Stars, Metal-Poor Stars, and the Galaxy, ed.
  A.~{McWilliam}, Vol.~5, 74

\bibitem[{{Kloppenborg} {et~al.}(2010){Kloppenborg}, {Stencel}, {Monnier},
  {Schaefer}, {Zhao}, {Baron}, {McAlister}, {ten Brummelaar}, {Che},
  {Farrington}, {Pedretti}, {Sallave-Goldfinger}, {Sturmann}, {Sturmann},
  {Thureau}, {Turner}, \& {Carroll}}]{Kloppenborg10}
{Kloppenborg}, B., {Stencel}, R., {Monnier}, J.~D., {et~al.} 2010, \nat, 464,
  870

\bibitem[{{Kopal}(1954)}]{Kopal54}
{Kopal}, Z. 1954, The Observatory, 74, 14

\bibitem[{{Laycock} {et~al.}(2010){Laycock}, {Tang}, {Grindlay}, {Los},
  {Simcoe}, \& {Mink}}]{Laycock10}
{Laycock}, S., {Tang}, S., {Grindlay}, J., {et~al.} 2010, \aj, 140, 1062

\bibitem[{{Lindegren} {et~al.}(2016){Lindegren}, {Lammers}, {Bastian},
  {Hern{\'a}ndez}, {Klioner}, {Hobbs}, {Bombrun}, {Michalik}, {Ramos-Lerate},
  {Butkevich}, {Comoretto}, {Joliet}, {Holl}, {Hutton}, {Parsons},
  {Steidelm{\"u}ller}, {Abbas}, {Altmann}, {Andrei}, {Anton}, {Bach},
  {Barache}, {Becciani}, {Berthier}, {Bianchi}, {Biermann}, {Bouquillon},
  {Bourda}, {Br{\"u}semeister}, {Bucciarelli}, {Busonero}, {Carlucci},
  {Casta{\~n}eda}, {Charlot}, {Clotet}, {Crosta}, {Davidson}, {de Felice},
  {Drimmel}, {Fabricius}, {Fienga}, {Figueras}, {Fraile}, {Gai}, {Garralda},
  {Geyer}, {Gonz{\'a}lez-Vidal}, {Guerra}, {Hambly}, {Hauser}, {Jordan},
  {Lattanzi}, {Lenhardt}, {Liao}, {L{\"o}ffler}, {McMillan}, {Mignard}, {Mora},
  {Morbidelli}, {Portell}, {Riva}, {Sarasso}, {Serraller}, {Siddiqui}, {Smart},
  {Spagna}, {Stampa}, {Steele}, {Taris}, {Torra}, {van Reeven}, {Vecchiato},
  {Zschocke}, {de Bruijne}, {Gracia}, {Raison}, {Lister}, {Marchant},
  {Messineo}, {Soffel}, {Osorio}, {de Torres}, \& {O'Mullane}}]{Lindegren16}
{Lindegren}, L., {Lammers}, U., {Bastian}, U., {et~al.} 2016, ArXiv e-prints,
  arXiv:1609.04303

\bibitem[{{Lisse} {et~al.}(2015){Lisse}, {Sitko}, \& {Marengo}}]{Lisse15}
{Lisse}, C.~M., {Sitko}, M.~L., \& {Marengo}, M. 2015, \apjl, 815, L27

\bibitem[{{Lund} {et~al.}(2016){Lund}, {Pepper}, {Stassun}, {Hippke}, \&
  {Angerhausen}}]{Lund16}
{Lund}, M.~B., {Pepper}, J., {Stassun}, K.~G., {Hippke}, M., \& {Angerhausen},
  D. 2016, ArXiv e-prints, arXiv:1605.02760

\bibitem[{{Magnier} {et~al.}(2013){Magnier}, {Schlafly}, {Finkbeiner}, {Juric},
  {Tonry}, {Burgett}, {Chambers}, {Flewelling}, {Kaiser}, {Kudritzki},
  {Morgan}, {Price}, {Sweeney}, \& {Stubbs}}]{Magnier13}
{Magnier}, E.~A., {Schlafly}, E., {Finkbeiner}, D., {et~al.} 2013, \apjs, 205,
  20

\bibitem[{{Marengo} {et~al.}(2015){Marengo}, {Hulsebus}, \&
  {Willis}}]{Marengo15}
{Marengo}, M., {Hulsebus}, A., \& {Willis}, S. 2015, \apjl, 814, L15

\bibitem[{{McQuillan} {et~al.}(2014){McQuillan}, {Mazeh}, \&
  {Aigrain}}]{Mcquillan14}
{McQuillan}, A., {Mazeh}, T., \& {Aigrain}, S. 2014, \apjs, 211, 24

\bibitem[{{Meibom} {et~al.}(2015){Meibom}, {Barnes}, {Platais}, {Gilliland},
  {Latham}, \& {Mathieu}}]{Meibom15}
{Meibom}, S., {Barnes}, S.~A., {Platais}, I., {et~al.} 2015, \nat, 517, 589

\bibitem[{{Miglio} {et~al.}(2012){Miglio}, {Brogaard}, {Stello}, {Chaplin},
  {D'Antona}, {Montalb{\'a}n}, {Basu}, {Bressan}, {Grundahl}, {Pinsonneault},
  {Serenelli}, {Elsworth}, {Hekker}, {Kallinger}, {Mosser}, {Ventura},
  {Bonanno}, {Noels}, {Silva Aguirre}, {Szabo}, {Li}, {McCauliff}, {Middour},
  \& {Kjeldsen}}]{miglio12}
{Miglio}, A., {Brogaard}, K., {Stello}, D., {et~al.} 2012, \mnras, 419, 2077

\bibitem[{{Moln{\'a}r} {et~al.}(2015){Moln{\'a}r}, {P{\'a}l}, {Plachy},
  {Ripepi}, {Moretti}, {Szab{\'o}}, \& {Kiss}}]{Molnar15}
{Moln{\'a}r}, L., {P{\'a}l}, A., {Plachy}, E., {et~al.} 2015, \apj, 812, 2

\bibitem[{{Mosser} {et~al.}(2012){Mosser}, {Goupil}, {Belkacem}, {Marques},
  {Beck}, {Bloemen}, {De Ridder}, {Barban}, {Deheuvels}, {Elsworth}, {Hekker},
  {Kallinger}, {Ouazzani}, {Pinsonneault}, {Samadi}, {Stello}, {Garc{\'{\i}}a},
  {Klaus}, {Li}, {Mathur}, \& {Morris}}]{Mosser12}
{Mosser}, B., {Goupil}, M.~J., {Belkacem}, K., {et~al.} 2012, \aap, 548, A10

\bibitem[{{Nemec} {et~al.}(2013){Nemec}, {Cohen}, {Ripepi}, {Derekas},
  {Moskalik}, {Sesar}, {Chadid}, \& {Bruntt}}]{Nemec13}
{Nemec}, J.~M., {Cohen}, J.~G., {Ripepi}, V., {et~al.} 2013, \apj, 773, 181

\bibitem[{{Quintana} {et~al.}(2010){Quintana}, {Jenkins}, {Clarke},
  {Chandrasekaran}, {Twicken}, {McCauliff}, {Cote}, {Klaus}, {Allen},
  {Caldwell}, \& {Bryson}}]{Quintana10}
{Quintana}, E.~V., {Jenkins}, J.~M., {Clarke}, B.~D., {et~al.} 2010, in
  \procspie, Vol. 7740, Software and Cyberinfrastructure for Astronomy, 77401X

\bibitem[{{Rappaport} {et~al.}(2014){Rappaport}, {Barclay}, {DeVore}, {Rowe},
  {Sanchis-Ojeda}, \& {Still}}]{Rappaport14}
{Rappaport}, S., {Barclay}, T., {DeVore}, J., {et~al.} 2014, \apj, 784, 40

\bibitem[{{Rappaport} {et~al.}(2012){Rappaport}, {Levine}, {Chiang}, {El
  Mellah}, {Jenkins}, {Kalomeni}, {Kite}, {Kotson}, {Nelson},
  {Rousseau-Nepton}, \& {Tran}}]{Rappaport12}
{Rappaport}, S., {Levine}, A., {Chiang}, E., {et~al.} 2012, \apj, 752, 1

\bibitem[{{Rodriguez} {et~al.}(2016){Rodriguez}, {Stassun}, {Lund}, {Siverd},
  {Pepper}, {Tang}, {Kafka}, {Gaudi}, {Conroy}, {Beatty}, {Stevens}, {Shappee},
  \& {Kochanek}}]{Rodriguez16}
{Rodriguez}, J.~E., {Stassun}, K.~G., {Lund}, M.~B., {et~al.} 2016, \aj, 151,
  123

\bibitem[{{Roeser} {et~al.}(2010){Roeser}, {Demleitner}, \&
  {Schilbach}}]{ppmxl}
{Roeser}, S., {Demleitner}, M., \& {Schilbach}, E. 2010, \aj, 139, 2440

\bibitem[{{Roettenbacher} {et~al.}(2016){Roettenbacher}, {Monnier}, {Korhonen},
  {Aarnio}, {Baron}, {Che}, {Harmon}, {K{\H o}v{\'a}ri}, {Kraus}, {Schaefer},
  {Torres}, {Zhao}, {Ten Brummelaar}, {Sturmann}, \&
  {Sturmann}}]{Roettenbacher16}
{Roettenbacher}, R.~M., {Monnier}, J.~D., {Korhonen}, H., {et~al.} 2016, \nat,
  533, 217

\bibitem[{{Sanchis-Ojeda} {et~al.}(2015){Sanchis-Ojeda}, {Rappaport},
  {Pall{\`e}}, {Delrez}, {DeVore}, {Gandolfi}, {Fukui}, {Ribas}, {Stassun},
  {Albrecht}, {Dai}, {Gaidos}, {Gillon}, {Hirano}, {Holman}, {Howard},
  {Isaacson}, {Jehin}, {Kuzuhara}, {Mann}, {Marcy}, {Miles-P{\'a}ez},
  {Monta{\~n}{\'e}s-Rodr{\'{\i}}guez}, {Murgas}, {Narita}, {Nowak}, {Onitsuka},
  {Paegert}, {Van Eylen}, {Winn}, \& {Yu}}]{SanchisOjeda15}
{Sanchis-Ojeda}, R., {Rappaport}, S., {Pall{\`e}}, E., {et~al.} 2015, \apj,
  812, 112

\bibitem[{{Schaefer}(2016)}]{Schaefer16}
{Schaefer}, B.~E. 2016, \apjl, 822, L34

\bibitem[{{Schuetz} {et~al.}(2016){Schuetz}, {Vakoch}, {Shostak}, \&
  {Richards}}]{Schuetz15}
{Schuetz}, M., {Vakoch}, D.~A., {Shostak}, S., \& {Richards}, J. 2016, \apjl,
  825, L5

\bibitem[{{Silva Aguirre} {et~al.}(2015){Silva Aguirre}, {Davies}, {Basu},
  {Christensen-Dalsgaard}, {Creevey}, {Metcalfe}, {Bedding}, {Casagrande},
  {Handberg}, {Lund}, {Nissen}, {Chaplin}, {Huber}, {Serenelli}, {Stello}, {Van
  Eylen}, {Campante}, {Elsworth}, {Gilliland}, {Hekker}, {Karoff}, {Kawaler},
  {Kjeldsen}, \& {Lundkvist}}]{SilvaAguirre15}
{Silva Aguirre}, V., {Davies}, G.~R., {Basu}, S., {et~al.} 2015, \mnras, 452,
  2127

\bibitem[{{Smith} {et~al.}(2012){Smith}, {Stumpe}, {Van Cleve}, {Jenkins},
  {Barclay}, {Fanelli}, {Girouard}, {Kolodziejczak}, {McCauliff}, {Morris}, \&
  {Twicken}}]{Smith12}
{Smith}, J.~C., {Stumpe}, M.~C., {Van Cleve}, J.~E., {et~al.} 2012, \pasp, 124,
  1000

\bibitem[{{Stellingwerf}(2013)}]{Stellingwerf13}
{Stellingwerf}, R.~F. 2013, ArXiv e-prints, arXiv:1310.0535

\bibitem[{{Strassmeier} \& {Rice}(2003)}]{Strassmeier03}
{Strassmeier}, K.~G., \& {Rice}, J.~B. 2003, \aap, 399, 315

\bibitem[{{Szab{\'o}} {et~al.}(2010){Szab{\'o}}, {Koll{\'a}th}, {Moln{\'a}r},
  {Kolenberg}, {Kurtz}, {Bryson}, {Benk{\H o}}, {Christensen-Dalsgaard},
  {Kjeldsen}, {Borucki}, {Koch}, {Twicken}, {Chadid}, {di Criscienzo}, {Jeon},
  {Moskalik}, {Nemec}, \& {Nuspl}}]{Szabo10}
{Szab{\'o}}, R., {Koll{\'a}th}, Z., {Moln{\'a}r}, L., {et~al.} 2010, \mnras,
  409, 1244

\bibitem[{{Tang} {et~al.}(2013{\natexlab{a}}){Tang}, {Grindlay}, {Los}, \&
  {Servillat}}]{Tang13a}
{Tang}, S., {Grindlay}, J., {Los}, E., \& {Servillat}, M. 2013{\natexlab{a}},
  \pasp, 125, 857

\bibitem[{{Tang} {et~al.}(2013{\natexlab{b}}){Tang}, {Sasselov}, {Grindlay},
  {Los}, \& {Servillat}}]{Tang13b}
{Tang}, S., {Sasselov}, D., {Grindlay}, J., {Los}, E., \& {Servillat}, M.
  2013{\natexlab{b}}, \pasp, 125, 793

\bibitem[{{Thompson} {et~al.}(2016){Thompson}, {Scicluna}, {Kemper}, {Geach},
  {Dunham}, {Morata}, {Ertel}, {Ho}, {Dempsey}, {Coulson}, {Petitpas}, \&
  {Kristensen}}]{Thompson16}
{Thompson}, M.~A., {Scicluna}, P., {Kemper}, F., {et~al.} 2016, \mnras, 458,
  L39

\bibitem[{{van Saders} {et~al.}(2016){van Saders}, {Ceillier}, {Metcalfe},
  {Silva Aguirre}, {Pinsonneault}, {Garc{\'{\i}}a}, {Mathur}, \&
  {Davies}}]{vanSaders16}
{van Saders}, J.~L., {Ceillier}, T., {Metcalfe}, T.~S., {et~al.} 2016, \nat,
  529, 181

\bibitem[{{Wright} {et~al.}(2016){Wright}, {Cartier}, {Zhao}, {Jontof-Hutter},
  \& {Ford}}]{Wright16}
{Wright}, J.~T., {Cartier}, K.~M.~S., {Zhao}, M., {Jontof-Hutter}, D., \&
  {Ford}, E.~B. 2016, \apj, 816, 17

\end{thebibliography}
\end{document}